\documentclass[11pt,a4paper]{article}
\pdfoutput=0
\usepackage{jcappub}

\usepackage[utf8]{inputenc}
\usepackage[T1]{fontenc}
\usepackage{graphicx,amssymb,amsmath,wasysym}
\usepackage{axodraw}
\usepackage{color}
\usepackage{multirow}

\newcommand{\vevf}{v_\phi}
\newcommand{\vevH}{v_H}
\newcommand{\lf}{\lambda_\phi}
\newcommand{\lfH}{\lambda_{\phi H}}
\newcommand{\lHX}{\lambda_{H X}}
\newcommand{\lfX}{\lambda_{\phi X}}
\newcommand{\lX}{\lambda_X}
\newcommand{\lH}{\lambda_H}
\newcommand{\lt}{\lambda_3}
\newcommand{\mh}{m_h}
\newcommand{\mr}{m_\rho}
\newcommand{\mx}{m_X}
\newcommand{\V}{{\cal V}}
\newcommand{\Z}{{\cal Z}}
\newcommand{\Zt}{{\mathbb{Z}_3}}

\begin{document}
\hfill {\tt ULB-TH/14-24}

	\title{WIMP and SIMP Dark Matter from the Spontaneous Breaking of a Global Group}

\author[1]{Nicolás Bernal,}
\emailAdd{nicolas@ift.unesp.br}

\author[2]{Camilo Garcia-Cely,}
\emailAdd{cgarciac@ulb.ac.de}

\author[1]{Rogério Rosenfeld}
\emailAdd{rosenfel@ift.unesp.br}

\affiliation[1]{ICTP South American Institute for Fundamental Research\\
                Instituto de Física Teórica, Universidade Estadual Paulista, São Paulo, Brazil}

\affiliation[2]{Service de Physique Théorique, Université Libre de Bruxelles,\\
                Boulevard du Triomphe, CP225, 1050 Brussels, Belgium}

\abstract{
We propose and study a scalar extension of the Standard Model which respects a $\Zt$ symmetry remnant of the spontaneous breaking of a global $U(1)_\text{DM}$ symmetry.
Consequently, this model has a natural dark matter candidate and a Goldstone boson in the physical spectrum.
In addition, the Higgs boson properties are changed with respect to the Standard Model
due to the mixing with a new particle.
We explore regions in the parameter space  taking into account bounds
from the measured Higgs properties, dark matter direct detection as well as measurements of the effective number of neutrino species before recombination. 
The dark matter relic density is determined by three classes of processes: the usual self-annihilation,
semi-annihilation and purely dark matter $3 \to 2$ processes. The latter has been subject of recent interest
leading to the so-called `Strongly Interacting Massive Particle' (SIMP) scenario. We show under which conditions
our model can lead to a concrete realization of
such scenario and study the possibility that the dark matter self-interactions could
address the small scale structure problems. In particular, we find that in order for the SIMP scenario to work,
the dark matter mass must be in the range $7-115$~MeV, with the global symmetry energy breaking scale in the TeV range. 
}

\maketitle

\section{Introduction}
The existence of dark matter (DM) comprising approximately $85 \%$ of the matter
content in the Universe, as supported by several astrophysical and cosmological observations,
is a strong evidence that  the highly successful Standard Model (SM) of particle physics is incomplete: there must be new physics beyond the Standard Model (BSM physics) 
(for a review, see {\it e.g.}~\cite{Jungman:1995df,Bergstrom:2000pn,Munoz:2003gx,Bertone:2004pz,2012AnP...524..479B}).

If DM is made out of new particles, these must be electrically neutral, have a lifetime longer than the
age of the Universe and be produced in the observed amount. 
There are several models containing particles that are suitable candidates for DM, such as supersymmetric (SUSY)
extensions of the SM with a discrete R-parity symmetry that makes the lightest supersymmetric particle
stable. However, at this point there is no evidence for this class of models from searches conducted at the Large Hadron Collider (LHC)
and the simplest SUSY models are under stress due to naturalness issues (see, {\it e.g.}~\cite{Giudice:2013yca}).

Another class of BSM scenarios with a DM candidate consists in extending the SM with an augmented scalar sector furnished with a discrete symmetry 
ensuring the  stability of the DM particle. If the extra scalars are postulated to be SM singlets,  they can interact with SM matter only through their couplings to the Higgs doublet, 
the so-called Higgs portal scenarios. 
This class of models may lead to invisible Higgs decays, allows for strongly self-interacting DM~\cite{Bento:2000ah,Bento:2001yk}
and may also alleviate the vacuum stability issue of the SM~\cite{Gonderinger:2009jp}. 

The simplest model in this class is the SM with an extra real scalar field respecting a discrete $\mathbb{Z}_2$ symmetry \cite{Silveira:1985rk,McDonald:1993ex,Burgess:2000yq}. 
Constraints on this scenario
have been recently updated~\cite{Cline:2013gha}. In fact, requiring the model to respect the observed DM abundance, 
direct detection bounds and  the limits on the invisible Higgs width 
severely constrains the parameters, placing the  extra scalar mass in the range between $53$ and $ 63$ GeV (see also~\cite{Feng:2014vea}). 

Another possibility is to consider scalar DM models with a $\mathbb{Z}_3$ symmetry.  One particular feature of these models is that they allow for semi-annihilation processes~\cite{Hambye:2008bq,D'Eramo:2010ep,Belanger:2012vp,Belanger:2014bga}, where the annihilation 
of two DM particles can give rise to a DM particle in the final state, as 
for example  $X X \rightarrow \bar{X} \Psi$, where $X$ and $\Psi$ stand for a DM and a non-DM particle respectively. 
In addition, they predict other interesting number-changing processes involving only DM 
particles, such as $XXX \to   X\bar{X}$. These inelastic processes can be relevant for the chemical
equilibrium and freeze-out of DM particles and their effect must be included in the Boltzmann equation controlling the
DM density. Remarkably, 
$\Zt$ models with that feature  
can also address the `core versus cusp'~\cite{Moore:1994yx,Flores:1994gz,Oh:2010mc,Walker:2011zu} and `too big to fail'~\cite{BoylanKolchin:2011de,Garrison-Kimmel:2014vqa} problems because  
they may exhibit sizable DM self-interactions. In this case, the DM in this type of scenarios has been
dubbed Strongly Interacting Massive Particle (SIMP)~\cite{Hochberg:2014dra}~\footnote{A model based on QCD-like strongly coupled gauge theories
with a Wess-Zumino-Witten term was recently put forward in~\cite{Hochberg:2014kqa}.}. 
$\Zt$ models have been recently studied, assuming that this symmetry is either just a global symmetry as in
\cite{Belanger:2012zr} or that it is a remnant of a local $U(1)_\text{DM}$ symmetry spontaneously broken to $\Zt$~\cite{Ko:2014nha}. 

In this paper we explore an intermediate situation, where a global $U(1)_\text{DM}$ symmetry is spontaneously broken to a discrete $\Zt$ symmetry by postulating the existence of a scalar field with  charge 3  under the global group and a non-vanishing vacuum expectation value~\cite{Krauss:1988zc}~\footnote{The  origin of the $U(1)_\text{DM}$ group could be attributed to an underlying flavor symmetry in order to generate radiative neutrino masses and to stabilize the DM as recently discussed in~\cite{Sierra:2014kua}.}. 
Due to this, in addition to a natural DM candidate there is a  Goldstone boson (GB) in the physical spectrum.
In our study, we include the novel processes mentioned above in the determination of the DM relic abundance and consider the role of GB in its production. In doing so we take into account bounds
from the measured Higgs properties, DM direct detection as well as measurements of the effective number of neutrino species before recombination. 
In addition, we find under which conditions
our model can be a concrete implementation of the SIMP scenario.

This paper is organized as follows. 
In the next section we introduce the model and discuss the different mechanisms that could generate the DM relic density.
In section~\ref{sec3} we analyze the constraints arising from the effective number of neutrino species, Higgs physics and direct detection experiments.
Following that, in section~\ref{sec:WIMP} we focus on the self- and semi-annihilating scenarios and show that in these cases DM behaves like a Weakly Interacting Massive Particle (WIMP).
Subsequently in section~\ref{sec:SIMP}, we address the possibility of having DM as a Strongly Interacting Massive Particle (SIMP).
In particular we study how to implement the $3\to2$ annihilation mechanism for its production and show that this naturally leads to self-interacting cross-sections that typically address small scale structure problems in astrophysics.
Finally, in section~\ref{sec:conclusions} we present our conclusions.

\section{$\boldsymbol{\Zt}$ Dark Matter from the Breaking of a Global $\boldsymbol{U(1)_\text{DM}}$}
\subsection{Description of the Model}
We postulate a dark sector with a global $U(1)_\text{DM}$ which is spontaneously broken into a $\Zt$ symmetry.
This is achieved by considering, in addition to the usual scalar doublet $H$, two complex scalar fields $\phi_X$ and $X$ which are 
singlets under the SM group.
While the SM particles do not transform under the $U(1)_\text{DM}$, the new scalars $\phi_X$ and $X$ have charges equal to $3$ and $1$ respectively.
This choice leads naturally to a $\Zt$ symmetry after the spontaneous breaking of the continuous group~\cite{Krauss:1988zc}.
In fact, if we assume that in the vacuum state $\langle X \rangle = 0$ and $\langle\phi_X \rangle\equiv\vevf\ne 0$, then 
the breaking $SU(2)_L\otimes U(1)_Y \otimes U(1)_\text{DM} \to U(1)_\text{EM} \otimes \Zt$ occurs.
To see this explicitly, let us consider the most general renormalizable scalar potential
\begin{eqnarray}
V=&-&\mu_H^2\,H^\dagger H+\lH(H^\dagger H)^2-\mu_\phi^2\,\phi_X^*\phi_X+\lf(\phi_X^*\phi_X)^2+\mu_X^2X^* X+\lX(X^* X)^2\nonumber\\
  &+&\lfH\phi_X^*\phi_X\,H^\dagger H+\lfX X^* X\phi_X^*\phi_X+\lHX X^* X\,H^\dagger H+\left(\lt X^3\phi_X^*+\text{H.c.}\right)\,.
\label{potential}
\end{eqnarray}
After symmetry breaking we can write the fields in the unitary gauge as
\begin{equation}
H = \frac{1}{\sqrt{2}}\left(\begin{array}{c}
0 \\ 
\vevH+\tilde{h}(x)
\end{array}  \right)\,,\qquad
\phi_X = \frac{1}{\sqrt{2}} \left(\vevf + \tilde{\rho} +i\,\eta \right),
\end{equation}
where $\vevH = 246$~GeV is the vacuum expectation value of $H$, which gives mass to the SM particles via the Brout-Englert-Higgs mechanism~\cite{Djouadi:2005gi}.
Therefore the part of the potential involving $X$ becomes 
\begin{eqnarray}
V_X =&&\mx^2 X^* X + \lX (X^* X)^2 + \frac12\lHX\left(2\vevH \tilde h + \tilde h^2\right) X^* X \nonumber\\
    &+& \frac12\lfX\left(2 \vevf\tilde\rho + \tilde\rho^2 + \eta^2\right) X^* X + \left( \frac{\lt}{\sqrt{2}} X^3 (\vevf + \tilde\rho - i\, \eta ) + \text{H.c.} \right)\,,
\label{VX}
\end{eqnarray}
where
\begin{equation}
\mx^2 = \mu_X^2 + \frac12 \lfX \vevf^2+ \frac12 \lHX \vevH^2\,.
\end{equation}

The potential in Eq.~\eqref{VX} is manifestly invariant under the remnant $\Zt\subset U(1)_\text{DM}$ that acts non-trivially only on the field $X$ in the following way: $X\to~e^{\frac{2\pi}{3}i}X$.
It is precisely because of this reason that the particle associated to $X$ can not decay and is thus identified with the DM.
In other words, the $\Zt$ group ensures the stability of the DM in the present model.
Notice also that while $X$ and $\bar X$ belong to different representations of $\Zt$, they have the same mass and therefore they both contribute to the total amount of DM.

Due to the $\lfH$ coupling there is a mixing between the $\tilde{h}$ and $\tilde{\rho}$ fields, which determines the masses for the physical states $h$ and $\rho$
\begin{eqnarray}
\mh^2 = \lH\,\vevH^2+\lf\,\vevf^2+\sqrt{(\lH\,\vevH^2-\lf\,\vevf^2)^2+(\lfH\,\vevH\,\vevf)^2}\,,\\
\mr^2 = \lH\,\vevH^2+\lf\,\vevf^2-\sqrt{(\lH\,\vevH^2-\lf\,\vevf^2)^2+(\lfH\,\vevH\,\vevf)^2}\,,
\end{eqnarray}
where we identify $h$ with the SM Higgs boson with mass $\mh\sim 125$~GeV and the mixing angle $\theta$ is defined by
\begin{equation}
\tan 2\theta=\frac{\lfH\,\vevH\,\vevf}{\lf\,\vevf^2-\lH\,\vevH^2}\,.
\end{equation}
From this it follows that the masses and $\vevf$ are related by
\begin{equation}
\vevf^2=\frac{\mh^2\sin^2\theta+\mr^2\,\cos^2\theta}{2\,\lf}\,.
\end{equation}
On the other hand, the field $\eta$ remains massless and it is hence the GB associated to the symmetry breaking of the global $U(1)_\text{DM}$.

Consequently, the physical spectrum of this model consists of the real scalars $h$, $\rho$ and the GB $\eta$, along with the complex scalar $X$.
There are ten free parameters in the potential.
Fixing $\mh$ and $\vevH$ to their physical values reduces the number of independent parameters to eight.
We will use the following parameters to characterize the model: two masses ($\mr$ and $\mx$), the mixing angle $\theta$ and five quartic couplings $\lf$, $\lX$, $\lt$, $\lfX$ and $\lHX$.
According to the potential~\eqref{potential}, these couplings are real with the exception of $\lt$, whose phase can be absorbed by a field redefinition, and therefore the dark sector does not introduce any $CP$-violation.

Let us point out that, with the exception of the term involving $\lt$, all the terms of Eq.~\eqref{VX} contain pairs of DM fields.
Hence, the $\lt$ coupling characterizes the $\Zt$ DM phenomenology.
On the other hand, $\lHX$ controls the so-called Higgs portal because it connects the SM with the DM particles via the Higgs boson.
Likewise, $\lfX$ connects the $\rho$ and $\eta$ bosons with the DM, and we therefore call it the GB portal.

Some comments are now in order.
Stability of the potential requires that it should be bounded from below for large values of the fields.
One could also require perturbativity of scattering amplitudes up to high energy scales, such as the Planck scale, which typically imply that the coupling constants should not be too large at the electroweak scale.
Some of these issues were discussed in~\cite{Ko:2014nha}.
For the purposes of this phenomenological analysis, we just assume that these conditions are met; however, we do require our couplings to be in the perturbative regime, $|\lambda| < 4 \pi$ at the electroweak scale.

The model was implemented in {\tt FeynRules}~\cite{Christensen:2008py,Christensen:2009jx} and the output was used both in {\tt CalcHEP}~\cite{Belyaev:2012qa} and {\tt MicrOMEGAs}~\cite{Belanger:2006is,Belanger:2013oya} in the phenomenological studies discussed in the following sections.

\subsection{Dark Matter Relic Abundance}
\label{sec:DMAbu}

Due to the $\Zt$ symmetry, in addition to the usual self-annihilation of $X$ and $\bar X$, there are other processes that change the number of DM particles, 
and hence contribute to the thermal abundance of DM.
For instance, two $X$ particles can semi-annihilate into a $\bar X$ and a $\Zt$ singlet state, or three $X$ particles can annihilate into a $X\bar X$ pair.
All these processes can be divided in three classes:
self-annihilation~\cite{Lee:1977ua}, semi-annihilation~\cite{Hambye:2008bq,D'Eramo:2010ep} and $3\to 2$ annihilations~\cite{Hochberg:2014dra}.
The first category corresponds to the usual processes where two DM particles
annihilate into non-DM particles. In addition, semi-annihilation occurs when
two DM particles produce a single DM particle in the final state.
Finally, $3\to 2$ processes involve only DM particles both in the initial and final states.

If we assume no asymmetry between $X$ and $\bar X$, the total DM number density $n$ associated to both particles satisfies the Boltzmann equation~\cite{Hambye:2009fg,Hochberg:2014dra}
\begin{equation}
\frac{dn}{dt}+3 H n=
- 2\left[ \left( \frac{n}{n_{\text{eq}}}\right)^2 -1 \right] \gamma_\text{self} -  
\frac{n}{n_{\text{eq}}}  \left[ \frac{n}{n_{\text{eq}}} -1 \right] \gamma_\text{semi} -
\left(\frac{n}{n_{\text{eq}}} \right)^2 \left[ \frac{n}{n_{\text{eq}}} -1 \right] \gamma_{3  \to 2}\,,
\label{be}
\end{equation}
where $H$ is the Hubble parameter and $n_\text{eq}$ is the equilibrium number density  for a given DM temperature $T$. The thermally-averaged interaction rates $\gamma$ are defined in Appendix~\ref{AppB}, and are given in terms of the cross-sections by
\begin{eqnarray}
&&\gamma_\text{self}=\frac{1}{2}n_\text{eq}^2\langle\sigma v\rangle_\text{self}\,,\qquad
\gamma_\text{semi}=n_\text{eq}^2\langle\sigma v\rangle_\text{semi}\,,\nonumber\\
&&\qquad\gamma_{3\to 2}=n_\text{eq}^2\langle\sigma v\rangle_{2\to 3}=n_\text{eq}^3\langle\sigma v^2\rangle_{3\to 2}\,.
\end{eqnarray}

The three terms on the right-hand-side of Eq.~\eqref{be} correspond to the self-annihilation, the semi-annihilation, and the $3\to 2$ annihilation processes, respectively.
Here $\langle\sigma v\rangle$ is the standard thermally-averaged velocity-weighted cross-section for $2\to n$ reactions and $\langle\sigma v^2\rangle$ 
is the equivalent quantity for $3\to n$ processes, as discussed in Appendix~\ref{AppB}.
Let us note that the Boltzmann equation should contain other terms corresponding to three DM particles annihilating into a two-body final state with none or one DM particle.
However these interactions are subdominant ({\it c.f.} Appendix~\ref{AppC}) and will be neglected in the following analysis.
Notice that $\langle \sigma v \rangle_{2 \to 3} = n_\text{eq}\,\langle \sigma v^2 \rangle_{3 \to 2}$ as shown in Appendix~\ref{AppB} using $CP$ conservation and the principle of the detailed balance~\cite{Gondolo:1990dk}.

In this work we solve the Boltzmann equation~\eqref{be} in order to estimate the DM relic density abundance today, which is given by
\begin{equation}
\Omega_\text{DM}=\left(2.742\cdot 10^{8}\,\text{GeV}^{-1}\right)\,\frac{n}{s}\,\mx\,,
\end{equation}
where $s$ is the entropy density of the photon plasma. That value must agree with the one measured by the Planck collaboration~\cite{Ade:2013zuv}:
\begin{equation}
\Omega_\text{DM}h^2=0.1196\pm0.0031\,.
\end{equation}

\begin{figure}[t]
\begin{center}
\includegraphics[width=8.7cm]{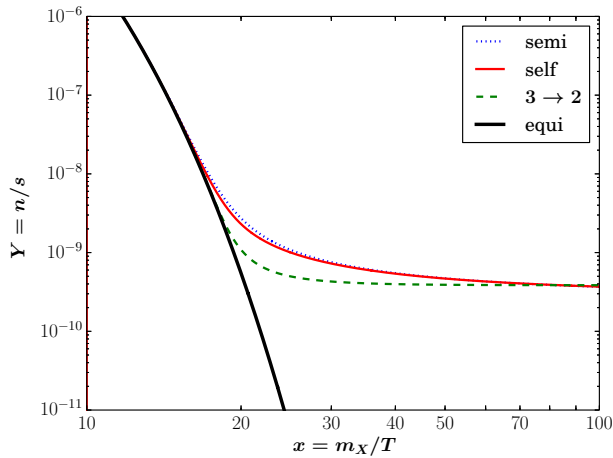}
\end{center}
\vspace{-0.8cm}
\caption{\sl \textbf{\textit{Freeze-out of the DM particles for the different thermal production modes,}}
for self-annihilation, semi-annihilation and the $3\to 2$ process, giving rise to the measured relic abundance (see text for details).
The equilibrium density is also depicted.
A DM mass $\mx=1$~GeV was assumed.
}
\label{figbe}
\end{figure}

For the sake of illustration, Fig.~\ref{figbe} shows the evolution of the comoving number densities of the DM as a function of $x\equiv\mx/T$ obtained by solving the Boltzmann equation for the different thermal production modes considered separately: self-annihilation (red), semi-annihilation (blue) and the $3\to 2$ process (orange). We take $\mx=1$~GeV with constant $\langle\sigma v\rangle_\text{self}$, $\langle\sigma v\rangle_\text{semi}$ and $\langle\sigma v^2\rangle_{3\to 2}$ chosen so that the final abundance equals the measured one. Since this plot is only for illustration, we assume that all the relativistic particles have the  same temperature (this might not always be the case, as shown in section~\ref{secdr}).
The corresponding equilibrium density is also depicted in black.
As shown in the figure, although the $3\to 2$ process freezes out at larger $x$ compared to the self- and semi-annihilations, the former reaches its final DM abundance much faster.

In sections~\ref{sec:WIMP} and~\ref{sec:SIMP}, we analyze further the different regions of parameter space where
the self-annihilation, semi-annihilation and $3\to 2$ processes are dominant and study the phenomenological consequences,
taking into account the constraints discussed in the following section.

\section{Constraints on the Model}\label{sec3}
In this section we discuss some constraints arising from the
production and decay of the SM-like Higgs, DM direct detection, as well as from the measurement of the effective number of neutrinos species before recombination.

\subsection{Effective Number of Neutrino Species}\label{secdr}

\begin{figure}[t]
\begin{center}
\includegraphics[scale=0.52]{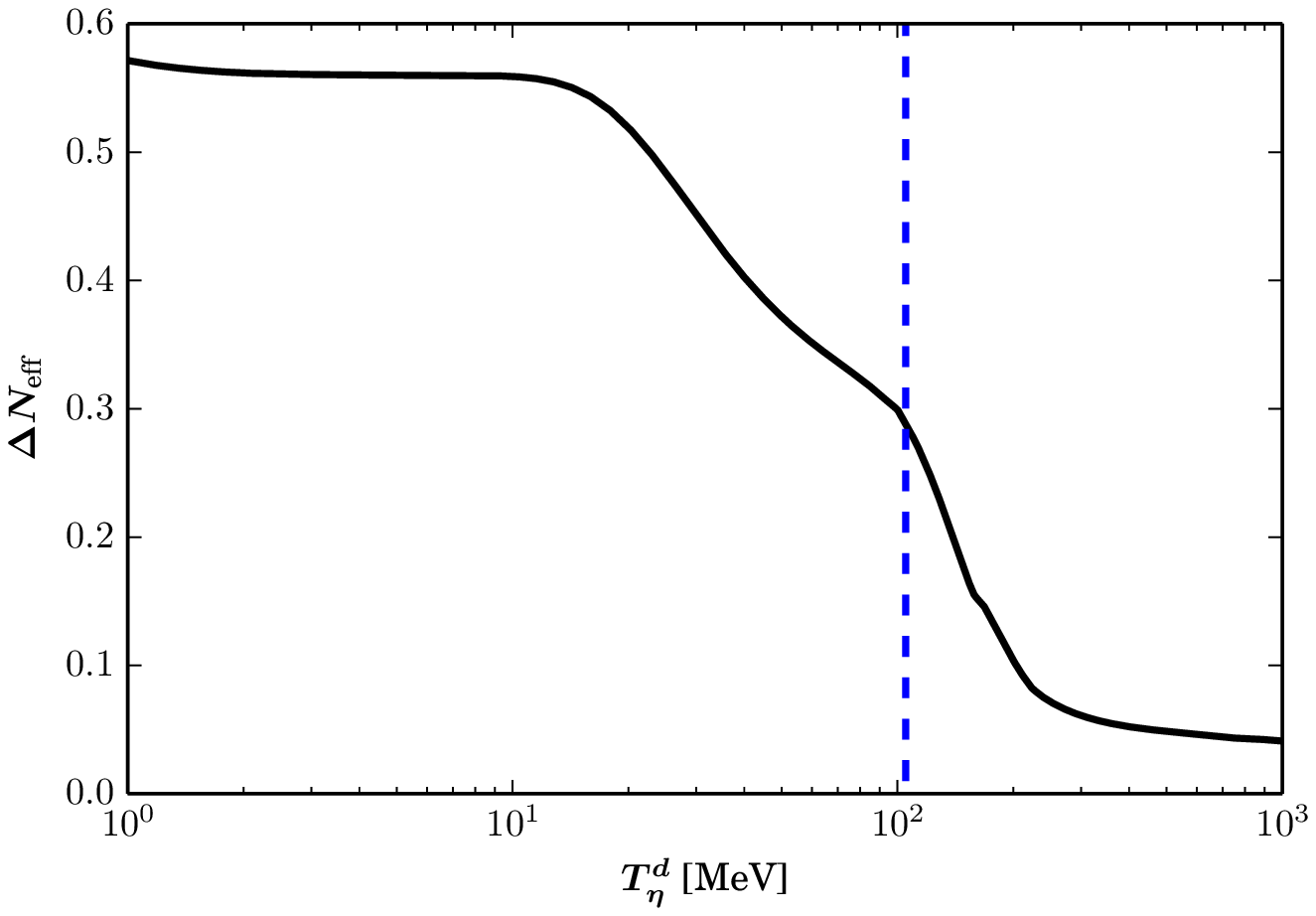}
\includegraphics[scale=0.58]{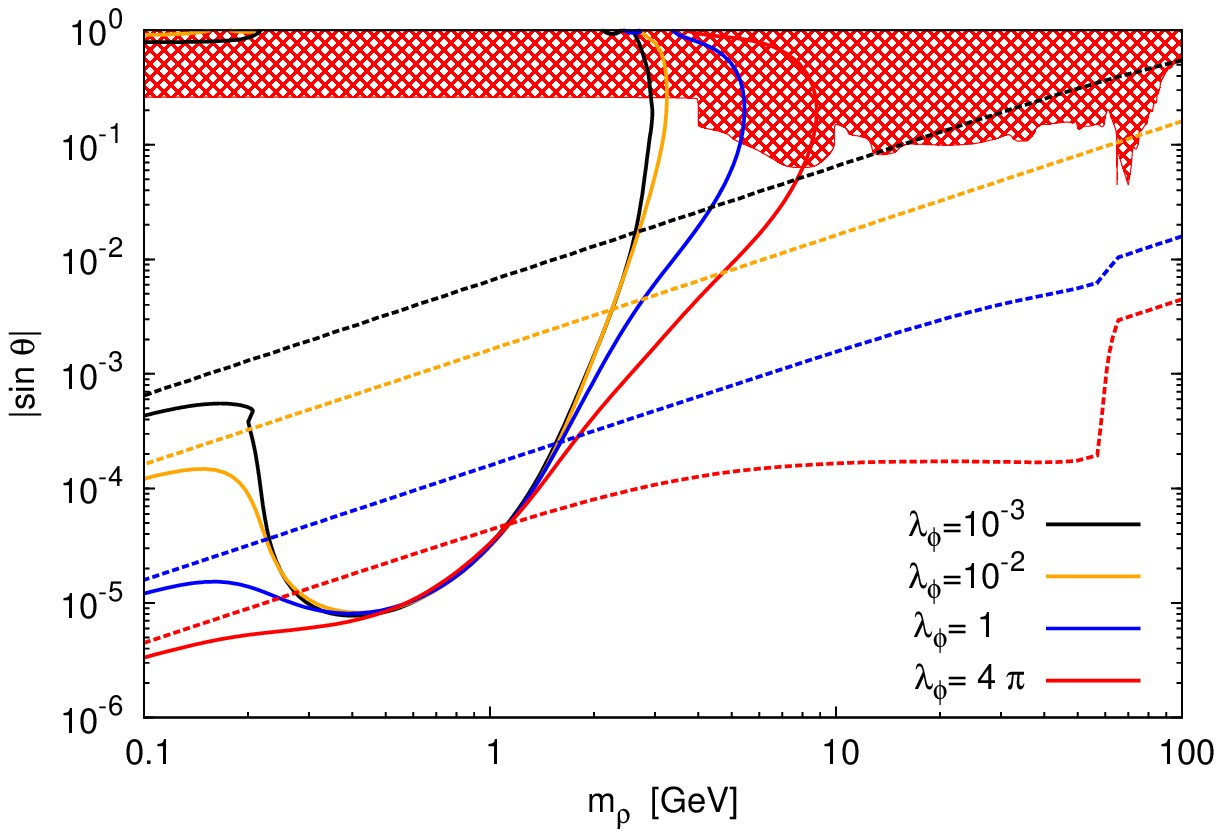}
\end{center}
\vspace{-0.8cm}
\caption{\sl 
\textbf{\textit{Left Panel.}}
GB contribution to the effective number of neutrino species  as a function of its decoupling temperature from the thermal plasma. The dashed vertical line corresponds to $T^d_\eta = m_\mu$.
\textbf{\textit{Right Panel.}} Mixing angle as a function of the $\rho$ boson mass. 
The continuous lines represent $T_\eta^d=m_\mu$.
The dashed lines correspond to the upper limit given by the invisible decay of the Higgs, assuming  $\mx>\mh/2$ so that the Higgs and the $\rho$ can not decay into a couple of DM particles. In addition, the hatched region shows the excluded mixing angles from $\rho$ searches in the OPAL detector at LEP.
}
\label{figDNeff}
\end{figure}
Being massless particles, GBs can fake the effects of neutrinos in the Cosmic Microwave Background (CMB).
The existence of these new relativistic degrees of freedom, the so-called Dark Radiation, 
is characterized by the effective number of neutrino species $N_\text{eff}$ (for a recent discussion see, {\it e.g.}~\cite{Steigman:2013yua}).
Measurements of the anisotropies in the CMB performed by the Planck collaboration in 2013 
indicate that $N_\text{eff} = 3.30 \pm 0.27$~\cite{Ade:2013zuv}, consistent with the SM prediction
\footnote{After the completion of this work, new preliminary results form the Planck collaboration indicate an even better agreement with the SM prediction $N_\text{eff} = 3.13 \pm 0.32$~\cite{Planck:2015xua}. These results are consistent with BBN~\cite{Steigman:12-2014}.}.
In this model, the contribution to the value of $N_\text{eff}$ from the GB is determined by the moment in which it 
goes out of equilibrium with the thermal plasma
\footnote{The equilibrium between the SM and the dark sector is achieved in the Early Universe by a number of processes.
For instance by $hh\to h\rho$, which is only suppressed by the mixing angle $\theta$, and which could establish the equilibrium at temperatures above the TeV scale.
We corroborate this for the scenarios considered in this work.}.
If $T^d_\eta$ and $T^d_\nu$ are respectively the decoupling temperatures of the GBs and neutrinos from the plasma and if the GBs are not reheated after their decoupling from the SM, then
\begin{equation}
	\Delta N_{\rm eff} = N_{\rm eff} -3 =
\frac{4}{7}\left[\frac{g(T^d_\nu)}{g(T^d_\eta)} \right]^{4/3},
\label{NeffdefG}
\end{equation}
where $g(T)$ is the number of relativistic degrees of freedom at a given temperature. We plot the resulting contribution as function of the decoupling temperature on the left panel of Fig.~\ref{figDNeff}. From this it is clear that when the decoupling temperature is greater than the muon mass, $N_\text{eff}$ is reasonably close to the SM prediction.  Consequently, we have to make sure that the interaction rates keeping the equilibrium between the GB and the SM become smaller than the Hubble expansion rate at temperatures greater than the muon mass.  In fact, assuming $\mr$ greater than a few GeV is enough to have a diluted GB contribution to $N_\text{eff}$.

\begin{figure}[t]
\begin{center}
{
\unitlength=1.0 pt
\SetScale{1.0}
\SetWidth{0.7}      
\scriptsize    
{} \qquad\allowbreak
\begin{picture}(96,38)(0,0)
\DashLine(36.0,23.0)(12.0,35.0){1.0}
\Text(12.0,35.0)[r]{$\eta$}
\DashLine(36.0,23.0)(12.0,11.0){1.0}
\Text(12.0,11.0)[r]{$\eta$}
\DashLine(36.0,23.0)(60.0,23.0){1.0}
\Text(49.0,24.0)[b]{$h,\,\rho$}
\ArrowLine(60.0,23.0)(84.0,35.0) 
\Text(84.0,35.0)[l]{$f$}
\ArrowLine(84.0,11.0)(60.0,23.0) 
\Text(84.0,11.0)[l]{$\bar f$}
\end{picture} \ 
{} \qquad\allowbreak
\begin{picture}(96,38)(0,0)
\DashLine(48.0,35.0)(24.0,35.0){1.0}
\Text(24.0,35.0)[r]{$\eta$}
\DashLine(48.0,35.0)(48.0,11.0){1.0}
\Text(49.0,24.0)[l]{$h,\,\rho$}
\DashLine(48.0,35.0)(72.0,35.0){1.0}
\Text(72.0,35.0)[l]{$\eta$}
\ArrowLine(24.0,11.0)(48.0,11.0)
\Text(24.0,11.0)[r]{$f$}
\ArrowLine(48.0,11.0)(72.0,11.0)
\Text(72.0,11.0)[l]{$f$}
\end{picture} \
}
\end{center}
\vspace{-0.8cm}
\caption{\sl Diagrams for the GB annihilation and scattering
relevant for the thermal decoupling.}
\label{figetaeta-mumu}
\end{figure}
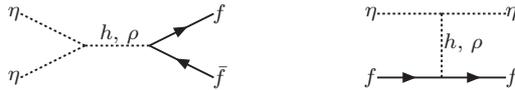

In order to show this, we consider the case $T^d_\eta \approx m_\mu$ (For a detailed analysis, see {\it e.g.}, \cite{Weinberg:2013kea,Garcia-Cely:2013nin}). Then the equilibrium takes place via reactions such as $\eta\,f\leftrightarrow\eta\,f$ and $f\,\bar f \leftrightarrow \eta\,\eta$, where $f$ denotes a SM fermion, as shown in Fig.~\ref{figetaeta-mumu}. Since the corresponding amplitudes are proportional to the fermion mass, processes involving muons are dominant.
Then, the departure of equilibrium occurs when the Hubble expansion 
becomes comparable to the interaction rate 
with muons.
Exploiting the fact that the latter depends only on the three parameters $\mr$, $\lf$ and $\theta$ (as well as on $T^d_\eta$), in Fig.~\ref{figDNeff}  we show the values of the mixing angle  for which the interaction rate equals the Hubble parameter for different values of $\lf$.
The regions below the solid lines correspond to decoupling temperatures  higher than the mass of the muon and thus to a diluted contribution to $N_\text{eff}$.
From Fig.~\ref{figDNeff} we conclude that such contribution is small when the $\rho$ scalar is heavier than 
about 5 GeV or when the mixing angle is smaller than about $10^{-5}$.

\subsection{Higgs Sector}

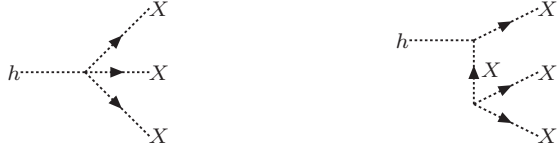
\begin{figure}[t]
\begin{center}
{
\unitlength=1.0 pt
\SetScale{1.0}
\SetWidth{0.7}      
\scriptsize    
{} \qquad\allowbreak
\begin{picture}(120,62)(0,0)
\DashLine(60.0,35.0)(36.0,35.0){1.0}
\Text(36.0,35.0)[r]{$h$}
\DashArrowLine(60.0,35.0)(84.0,59.0){1.0} 
\Text(84.0,59.0)[l]{$X$}
\Text(84.0,62.0)[l]{$~$}
\DashArrowLine(60.0,35.0)(84.0,35.0){1.0} 
\Text(84.0,35.0)[l]{$X$}
\Text(84.0,38.0)[l]{$~$}
\DashArrowLine(60.0,35.0)(84.0,11.0){1.0} 
\Text(84.0,11.0)[l]{$X$}
\Text(84.0,14.0)[l]{$~$}
\end{picture} \ 
{} \qquad\allowbreak
\begin{picture}(120,62)(0,0)
\DashLine(60.0,47.0)(36.0,47.0){1.0}
\Text(36.0,47.0)[r]{$h$}
\DashArrowLine(60.0,47.0)(84.0,59.0){1.0} 
\Text(84.0,59.0)[l]{$X$}
\Text(84.0,62.0)[l]{$~$}
\DashArrowLine(60.0,23.0)(60.0,47.0){1.0} 
\Text(63.0,36.0)[l]{$X$}
\Text(63.0,39.0)[l]{$~$}
\DashArrowLine(60.0,23.0)(84.0,35.0){1.0} 
\Text(84.0,35.0)[l]{$X$}
\Text(84.0,38.0)[l]{$~$}
\DashArrowLine(60.0,23.0)(84.0,11.0){1.0} 
\Text(84.0,11.0)[l]{$X$}
\Text(84.0,14.0)[l]{$~$}
\end{picture} \ 
}
\end{center}
\vspace{-0.8cm}
\caption{\sl \textbf{\textit{Tree level diagrams for the decay $\boldsymbol{h \to X X X}$.}}
}
\label{hxxx}
\end{figure}

The enlarged scalar sector increases the number of Higgs decay channels.
The new decay modes are into pairs of $\rho$, $\eta$ and DM particles, when kinematically allowed.
The last two channels ($h\to\eta\,\eta$ and $h\to X\,\bar{X}$) contribute to the invisible decay of the Higgs, whereas the first one ($h\to\rho\,\rho$) only does 
it partially when the $\rho$ pair itself decays into a couple of $\eta$ or DM particles.
The corresponding decay rates are:
\begin{eqnarray}
\Gamma(h\to\eta\,\eta)&=&\frac{\mh^3\,\sin^2\theta}{32\pi\,\vevf^2}\,,\\
\Gamma(h\to\rho\,\rho)&=&\frac{(\mh^2+2\,\mr^2)^2}{128\pi\,\mh^2\,\vevH^2\,\vevf^2}\,\sqrt{\mh^2-4\mr^2}\,\left(\vevH\,\cos\theta-\vevf\,\sin\theta\right)^2\,\sin^22\theta\,,\\
\Gamma(h\to X\,\bar{X})&=&\frac{\sqrt{\mh^2-4\mx^2}}{32\pi\,\mh^2}\,\left(\lfX\,\vevf\,\sin\theta-\lHX\,\vevH\,\cos\theta\right)^2\,.
\end{eqnarray}
Notice that three DM particles form a $\Zt$-singlet and as a consequence the decays $h~\rightarrow~XXX$ and $h~\rightarrow\bar X\bar X\bar X$ are allowed.
The corresponding Feynman diagrams are shown in Fig.~\ref{hxxx}, where one can see that its rate is proportional to $\lt^2$.
Nevertheless, this channel is phase-space suppressed and therefore subdominant compared to the two-body decay modes.

Correspondingly, the decay channels for the $\rho$ boson are
\begin{eqnarray}
\Gamma(\rho\to\eta\,\eta)&=&\frac{\mr^3\,\cos^2\theta}{32\pi\,\vevf^2}\,,\\
\Gamma(\rho\to X\,\bar{X})&=&\frac{\sqrt{\mr^2-4\mx^2}}{32\pi\,\mr^2}\left(\lfX\,\vevf\,\cos\theta+\lHX\,\vevH\,\sin\theta\right)^2\,.
\end{eqnarray}

In addition, due to the $\tilde{h}-\tilde{\rho}$ mixing, the $\rho$ boson can also decay into SM particles, with a rate give by $\Gamma(\rho\to\text{SM SM})=\Gamma(h_\text{SM}\to\text{SM SM})\times\sin^2\theta$, taking $m_{h_\text{SM}} \to \mr$ in the right-hand-side of the expression.
Because of the same reason, all the decay widths of the $h$ boson into SM particles decrease compared to the SM ones by a factor $\cos^2\theta$.

With these decay rates, we calculate the invisible Higgs branching ratio and apply the bound reported in~\cite{Bechtle:2014ewa}. 
That analysis considers simultaneously a universal modification of the Higgs boson couplings to SM particles as well as the possibility of an invisible Higgs decay,  concluding that
\begin{equation}
\kappa^2  \left[ 1 - \text{BR}_\text{inv} \right] \geq 0.81 \;\;\; (@\;95 \%\; \mbox{C.L.}),
\label{bound}
\end{equation}
where $\kappa$ is the universal modification for the Higgs coupling.
In our case we identify $\kappa=\cos\theta$ and the invisible Higgs decay branching ratio with
\begin{eqnarray}
\text{BR}_\text{inv} = \text{BR}(h\to\eta\,\eta)+\text{BR}(h\to X\,\bar{X})+\text{BR}(h\to X\,X\,X)+\text{BR}(h\to \bar X\,\bar X\,\bar X)\nonumber\\
+\text{BR}(h\to\rho\,\rho)\times\left[\text{BR}(\rho\to\eta\,\eta)+\text{BR}(\rho\to X\,\bar{X})\right]^2\,.
\end{eqnarray}

Although this is the most stringent constraint, we also require that the total decay width never exceeds the experimental bound~\cite{Khachatryan:2014iha}
\begin{equation}
\Gamma_h^\text{tot}\lesssim 22~\text{MeV} \;\;\; (@\;95 \%\; \mbox{C.L.})\,.
\end{equation}

Direct searches for a light neutral scalar produced in association with a $Z$ in the OPAL detector~\cite{Abbiendi:2002qp} at LEP also apply.
By identifying such a particle with the $\rho$ scalar, it is possible to set an upper bound on the mixing angle.

We apply these limits for all the scenarios considered in this work.
In the right panel of Fig.~\ref{figDNeff} we show them in the particular case when the scalars can not decay into DM particles.
The dashed lines represent the upper bounds arising from the Higgs invisible decay of Eq.~\eqref{bound}, whereas the upper hatched region is excluded because of direct searches of the $\rho$ scalar in OPAL.

\subsection{Dark Matter Direct Detection}

\begin{figure}[t]
\begin{center}
{
\unitlength=1.0 pt
\SetScale{1.0}
\SetWidth{0.7}      
\scriptsize    
{} \qquad\allowbreak
\begin{picture}(96,38)(0,0)
\DashArrowLine(24.0,35.0)(48.0,35.0){1.0} 
\Text(24.0,35.0)[r]{$X$}
\DashLine(48.0,35.0)(48.0,11.0){1.0}
\Text(49.0,24.0)[l]{$h$}
\DashArrowLine(48.0,35.0)(72.0,35.0){1.0} 
\Text(72.0,35.0)[l]{$X$}
\Text(72.0,38.0)[l]{$~$}
\ArrowLine(24.0,11.0)(48.0,11.0) 
\Text(24.0,11.0)[r]{$q$}
\ArrowLine(48.0,11.0)(72.0,11.0) 
\Text(72.0,11.0)[l]{$q$}
\end{picture} \ 
{} \qquad\allowbreak
\begin{picture}(96,38)(0,0)
\DashArrowLine(24.0,35.0)(48.0,35.0){1.0} 
\Text(24.0,35.0)[r]{$X$}
\DashLine(48.0,35.0)(48.0,11.0){1.0}
\Text(49.0,24.0)[l]{$\rho$}
\DashArrowLine(48.0,35.0)(72.0,35.0){1.0} 
\Text(72.0,35.0)[l]{$X$}
\Text(72.0,38.0)[l]{$~$}
\ArrowLine(24.0,11.0)(48.0,11.0) 
\Text(24.0,11.0)[r]{$q$}
\ArrowLine(48.0,11.0)(72.0,11.0) 
\Text(72.0,11.0)[l]{$q$}
\end{picture} \ 
}
\end{center}
\vspace{-0.8cm}
\caption{\sl \textbf{\textit{Diagrams responsible for DM direct detection.}}
}
\label{figxq-xq}
\end{figure}
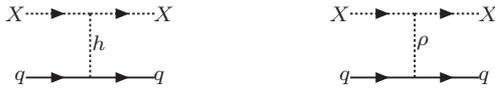

The scattering of DM particles off nuclei in direct detection experiments takes place via the $t$-channel exchange of $h$ and $\rho$ bosons
as depicted in Fig.~\ref{figxq-xq}.
Such process only generates a spin-independent signal with a DM-nucleon cross-section given by
\begin{equation}
\sigma_{XN} = \frac{ f_N^2\,  m_N^4 \left[ \lHX\,\vevH (\mh^2 + \mr^2) +(\lHX\,\vevH \cos2\theta -\lfX\,\vevf \sin2\theta )(\mr^2-\mh^2 )    \right]^2 }{4\pi \,\vevH^2\,\mh^4\,\mr^4\,(\mx+m_N)^2 }\,,
\label{DirectDetectionEq}
\end{equation}
where $m_N$ denotes the nucleon mass and $f_N \approx 0.27$ is a constant that depends on the nucleon matrix element~\cite{Belanger:2013oya}. 
For GeV Dark Matter, the resulting expression is constrained by  comparing it to the upper bound obtained by the LUX collaboration~\cite{Akerib:2013tjd}.

\section{Weakly Interacting Dark Matter }
\label{sec:WIMP}

If self- or semi-annihilations dominate the production of DM, the latter is  in kinetic equilibrium with the SM particles and the GB (See Appendix~\ref{AppB}). As discussed in section \ref{secdr}, due to the constraint on $N_\text{eff}$,  this equilibrium can only exist at temperatures greater than the muon mass. Since the freeze-out temperature is roughly $m_X/25 $, we thus consider only self- and semi-annihilations in the GeV range.    

Because of this mass range and since the couplings involved in DM phenomenology are typically small, self and semi-annihilation scenarios describe Weakly Interacting Massive Particles (WIMP).  We now discuss them separately.

\subsection{Self-annihilating Scenario}
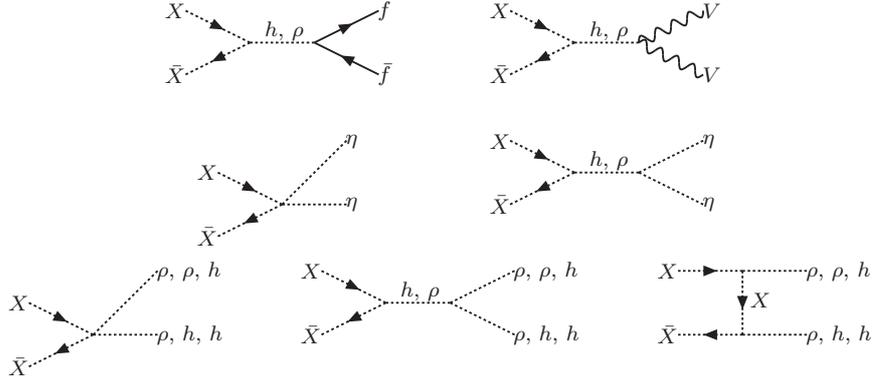
\begin{figure}[t]
\begin{center}
{
\unitlength=1.0 pt
\SetScale{1.0}
\SetWidth{0.7}      
\scriptsize    
{} \qquad\allowbreak
\begin{picture}(96,38)(0,0)
\DashArrowLine(12.0,35.0)(36.0,23.0){1.0} 
\Text(12.0,35.0)[r]{$~X$}
\DashArrowLine(36.0,23.0)(12.0,11.0){1.0} 
\Text(12.0,11.0)[r]{$~\bar X$}
\DashLine(36.0,23.0)(60.0,23.0){1.0}
\Text(49.0,24.0)[b]{$h,\,\rho$}
\ArrowLine(60.0,23.0)(84.0,35.0) 
\Text(84.0,35.0)[l]{$f$}
\ArrowLine(84.0,11.0)(60.0,23.0) 
\Text(84.0,11.0)[l]{$\bar f$}
\end{picture} \ 
{} \qquad\allowbreak
\begin{picture}(96,38)(0,0)
\DashArrowLine(12.0,35.0)(36.0,23.0){1.0}
\Text(12.0,35.0)[r]{$~X$}
\DashArrowLine(36.0,23.0)(12.0,11.0){1.0}
\Text(12.0,11.0)[r]{$~\bar X$}
\DashLine(36.0,23.0)(60.0,23.0){1.0}
\Text(49.0,24.0)[b]{$h,\,\rho$}
\Photon(60.0,23.0)(84.0,35.0){2.0}{4.0}
\Text(84.0,35.0)[l]{$V$}
\Photon(84.0,11.0)(60.0,23.0){2.0}{4.0}
\Text(84.0,11.0)[l]{$V$}
\end{picture} \
}
\vspace{-0.6cm}
{
\unitlength=1.0 pt
\SetScale{1.0}
\SetWidth{0.7}      
\scriptsize    
{} \qquad\allowbreak
\begin{picture}(96,38)(0,0)
\DashArrowLine(24.0,23.0)(48.0,11.0){1.0} 
\Text(24.0,23.0)[r]{$~X$}
\DashArrowLine(48.0,11.0)(24.0,-1.0){1.0} 
\Text(24.0,-1.0)[r]{$~\bar X$}
\DashLine(48.0,11.0)(72.0,35.0){1.0}
\Text(72.0,35.0)[l]{$\eta$}
\DashLine(48.0,11.0)(72.0,11.0){1.0}
\Text(72.0,11.0)[l]{$\eta$}
\end{picture} \ 
{} \qquad\allowbreak
\begin{picture}(96,38)(0,0)
\DashArrowLine(12.0,35.0)(36.0,23.0){1.0} 
\Text(12.0,35.0)[r]{$~X$}
\DashArrowLine(36.0,23.0)(12.0,11.0){1.0} 
\Text(12.0,11.0)[r]{$~\bar X$}
\DashLine(36.0,23.0)(60.0,23.0){1.0}
\Text(49.0,24.0)[b]{$h,\,\rho$}
\DashLine(60.0,23.0)(84.0,35.0){1.0}
\Text(84.0,35.0)[l]{$\eta$}
\DashLine(60.0,23.0)(84.0,11.0){1.0}
\Text(84.0,11.0)[l]{$\eta$}
\end{picture} \ 
}
\vspace{-0.6cm}
{
\unitlength=1.0 pt
\SetScale{1.0}
\SetWidth{0.7}      
\scriptsize    
{} \qquad\allowbreak
\begin{picture}(96,38)(0,0)
\DashArrowLine(24.0,23.0)(48.0,11.0){1.0} 
\Text(24.0,23.0)[r]{$~X$}
\DashArrowLine(48.0,11.0)(24.0,-1.0){1.0} 
\Text(24.0,-1.0)[r]{$~\bar X$}
\DashLine(48.0,11.0)(72.0,35.0){1.0}
\Text(72.0,35.0)[l]{$\rho,\,\rho,\,h$}
\DashLine(48.0,11.0)(72.0,11.0){1.0}
\Text(72.0,11.0)[l]{$\rho,\,h,\,h$}
\end{picture} \ 
{} \qquad\allowbreak
\begin{picture}(96,38)(0,0)
\DashArrowLine(12.0,35.0)(36.0,23.0){1.0} 
\Text(12.0,35.0)[r]{$~X$}
\DashArrowLine(36.0,23.0)(12.0,11.0){1.0} 
\Text(12.0,11.0)[r]{$~\bar X$}
\DashLine(36.0,23.0)(60.0,23.0){1.0}
\Text(49.0,24.0)[b]{$h,\,\rho$}
\DashLine(60.0,23.0)(84.0,35.0){1.0}
\Text(84.0,35.0)[l]{$\rho,\,\rho,\,h$}
\DashLine(60.0,23.0)(84.0,11.0){1.0}
\Text(84.0,11.0)[l]{$\rho,\,h,\,h$}
\end{picture} \ 
{} \qquad\allowbreak
\begin{picture}(96,38)(0,0)
\DashArrowLine(24.0,35.0)(48.0,35.0){1.0} 
\Text(24.0,35.0)[r]{$~X$}
\DashArrowLine(48.0,35.0)(48.0,11.0){1.0} 
\Text(51.0,24.0)[l]{$X$}
\Text(51.0,27.0)[l]{$~$}
\DashLine(48.0,35.0)(72.0,35.0){1.0}
\Text(72.0,35.0)[l]{$\rho,\,\rho,\,h$}
\DashArrowLine(48.0,11.0)(24.0,11.0){1.0} 
\Text(24.0,11.0)[r]{$~\bar X$}
\DashLine(48.0,11.0)(72.0,11.0){1.0}
\Text(72.0,11.0)[l]{$\rho,\,h,\,h$}
\end{picture} \ 
{} \qquad\allowbreak
}
\end{center}
\vspace{-0.8cm}
\caption{\sl \textbf{\textit{Tree level diagrams for the annihilation between $\boldsymbol{X}$ and $\boldsymbol{\bar X}$}.}
Here $f$ corresponds to any SM fermion and $V$ to any massive gauge boson.}
\label{selfann}
\end{figure}
If the self-annihilation of DM dominates over the other two processes, that is if $\gamma_\text{self}$ is much greater than $\gamma_\text{semi}$ and $\gamma_{3\to 2}$, the DM production proceeds via the familiar Lee-Weinberg scenario~\cite{Lee:1977ua}.
In this case, as shown in Fig.~\ref{selfann}, the DM abundance is obtained by self-annihilation into $\eta\,\eta$, $\rho\,\rho$, $h\,\rho$ and pairs of SM particles, until the reactions freeze-out at $x\sim 25$.
DM annihilates into SM fermions and vector bosons via the $s$-channel exchange of a $h$ or a $\rho$ boson.
For the other annihilation channels the contact term and the $t$-channel exchange of a $X$ are also present.
The only exception is the reaction $X\bar X\to\eta\eta$ where the $t$-channel exchange does not exist.
All these diagrams are proportional to $\lHX$ and $\lfX$ as opposed to the ones corresponding to the semi-annihilation and the $3\to 2$ mechanisms which are proportional to $\lt$.
As a result, self-annihilation dominates whenever $\lt$ is much smaller than $\lHX$ and $\lfX$.

\begin{figure}[t]
\begin{center}
\includegraphics[width=8.7cm]{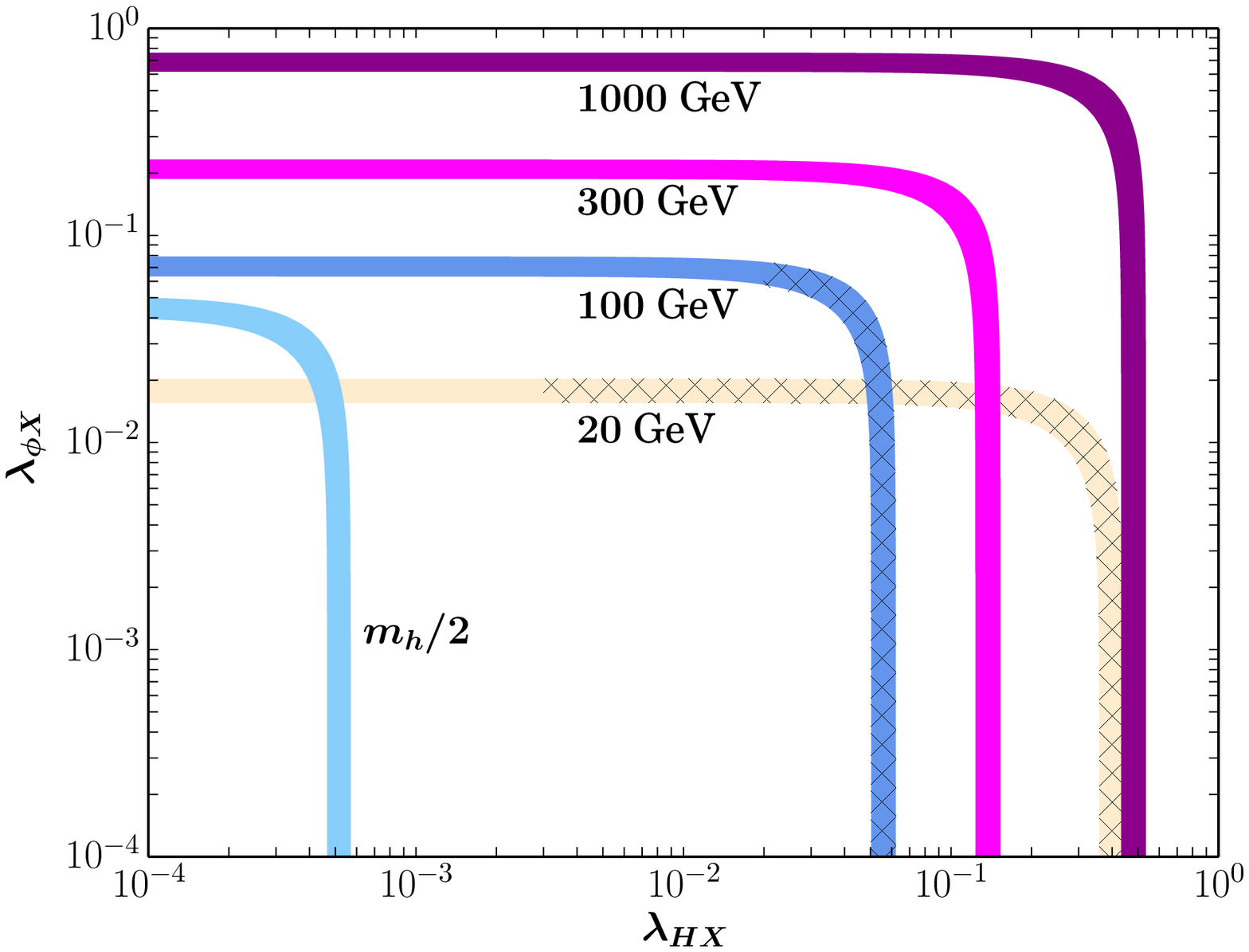}
\includegraphics[width=7.6cm]{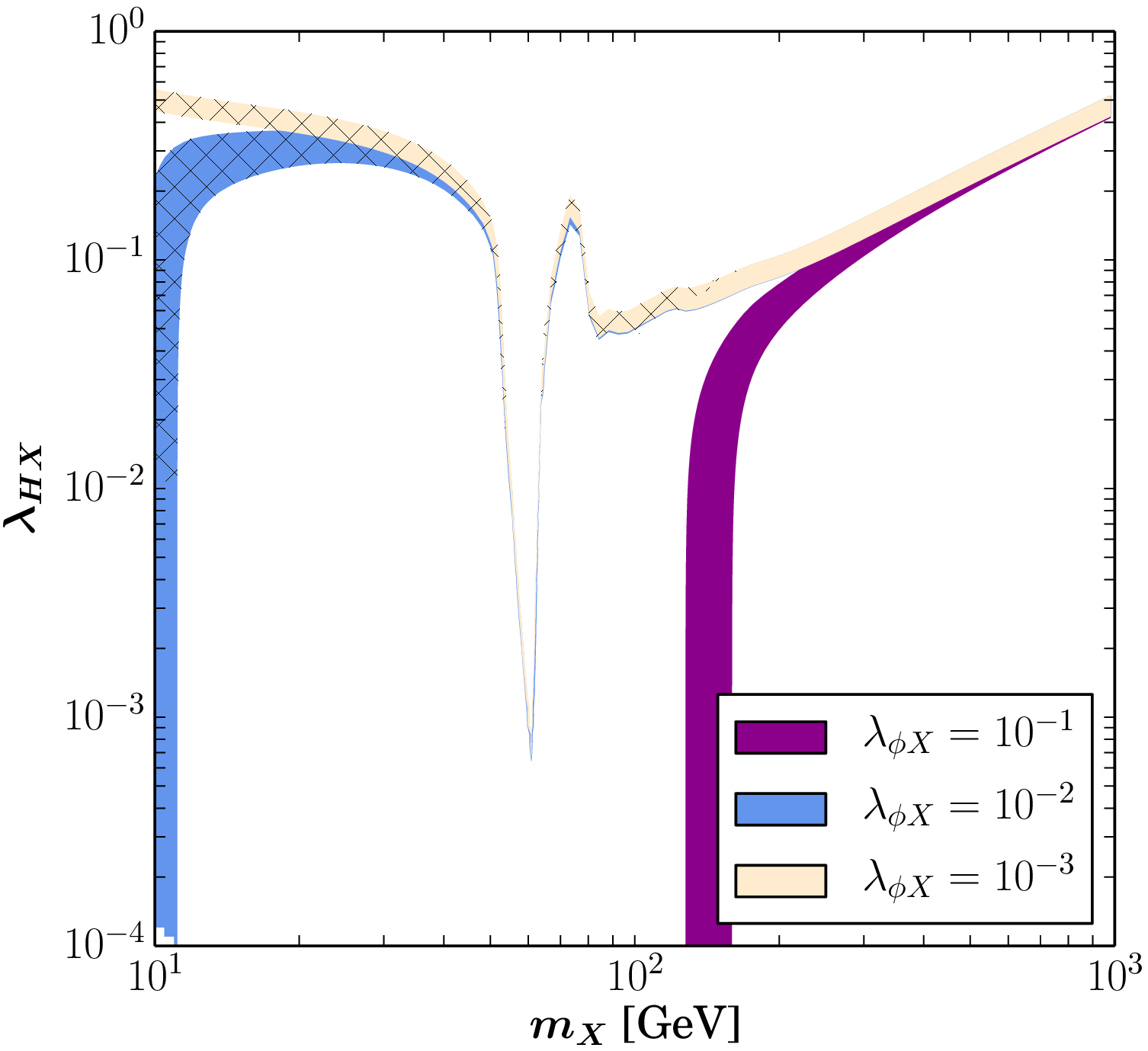}
\includegraphics[width=7.6cm]{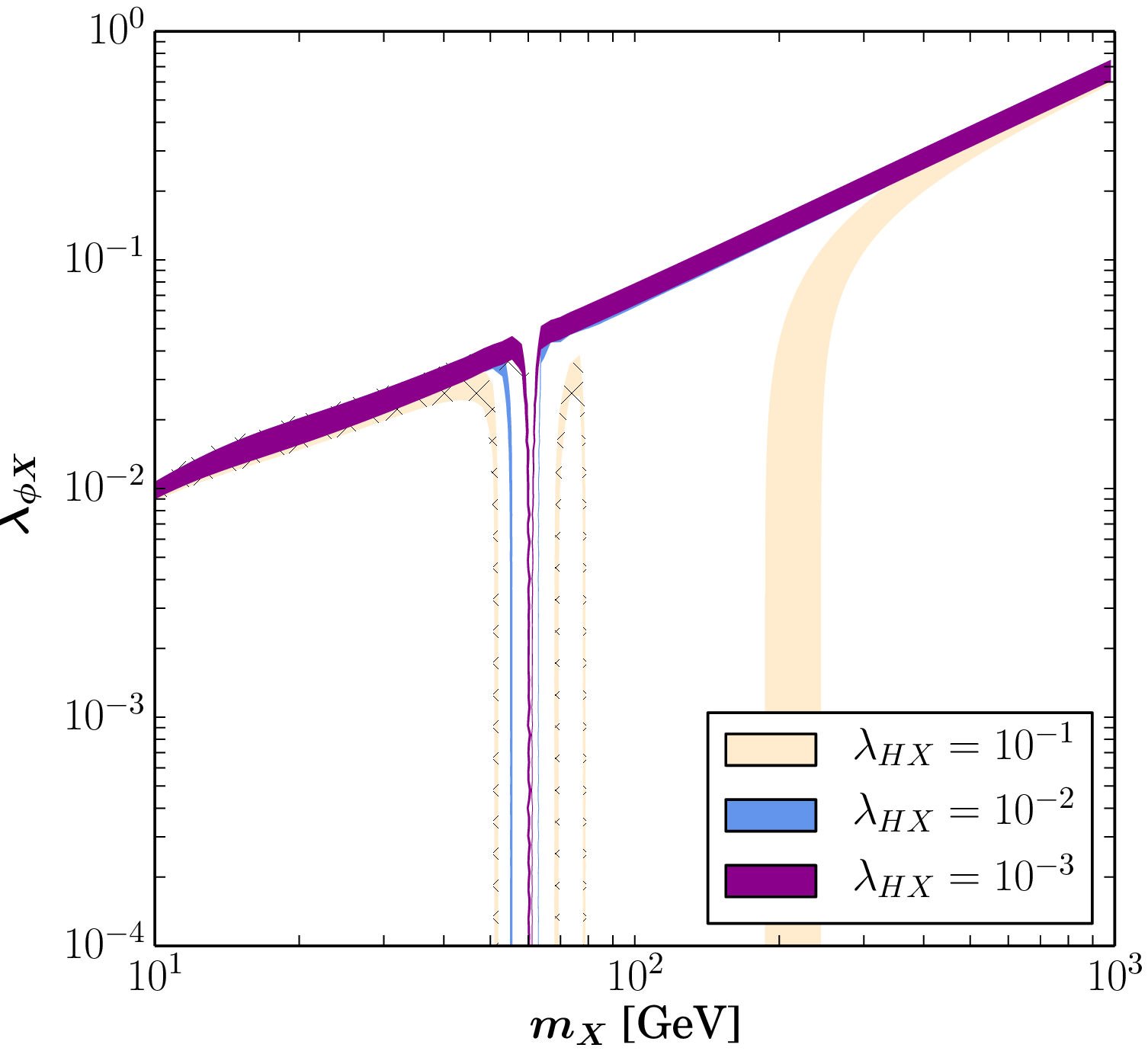}
\end{center}
\vspace{-0.8cm}
\caption{\sl \textbf{\textit{Self-annihilation scenario.}}
Parameter space giving rise to the measured DM relic abundance, for different combinations of $\mx$ and the Higgs and GB portals $\lHX$ and $\lfX$.
We have chosen $\mr=5$~GeV, $\sin\theta=10^{-5}$ and $\lf=0.1$, we also set $\lt=0$ in order to avoid semi-annihilations and $3\to 2$ reactions.
The hatched regions are ruled out by the LUX experiment.
}
\label{figdd}
\end{figure}
If we neglect the mixing angle, there are two regimes depending on the relative size of the Higgs and GB portals $\lHX$ and $\lfX$.
On the one hand, when the latter dominates, DM annihilates into the dark sector, that is, into GB and $\rho$ pairs.
The two channels are democratically distributed provided that $\mx\gg\mr$.
On the other hand, when $\lHX\gg\lfX$, the annihilation produces SM particles pairs.
Due to the $s$-channel Higgs exchange, the heaviest kinematically allowed pair tend to be produced.
These two regimes are illustrated in the upper panel of Fig.~\ref{figdd}, where we show the parameter space giving rise to the measured DM relic abundance, for different combinations of $\mx$, $\lHX$ and $\lfX$.
There we take $\mr=5$~GeV, $\sin\theta=10^{-5}$, $\lf=0.1$ as well as $\lt=0$ in order to avoid semi-annihilations  and $3\to 2$ annihilations. We calculate the relic abundance using {\tt MicrOMEGAs-3}.
From the figure it is possible to see that large masses in general require large couplings.
We can also see that when $\lHX$ dominates, due to the existence of the Higgs resonance and kinematic thresholds, the dependence on the mass is more complicated than in the other case.
This is exemplified in the lower panels of Fig.~\ref{figdd}, where the dependence on the DM mass is shown explicitly.
The Higgs boson funnel appears as well as the kinematic openings of the annihilations into $W^+\,W^-$, $Z\,Z$ and $h\,h$.
The LUX experiment rules out parts of the parameter space corresponding to DM masses $\mathcal{O}\left(10\right)$~GeV and sizable $\lHX$ couplings.
This corresponds to the hatched regions in Fig.~\ref{figdd}.

\begin{figure}[t]
\begin{center}
\includegraphics[width=7.6cm]{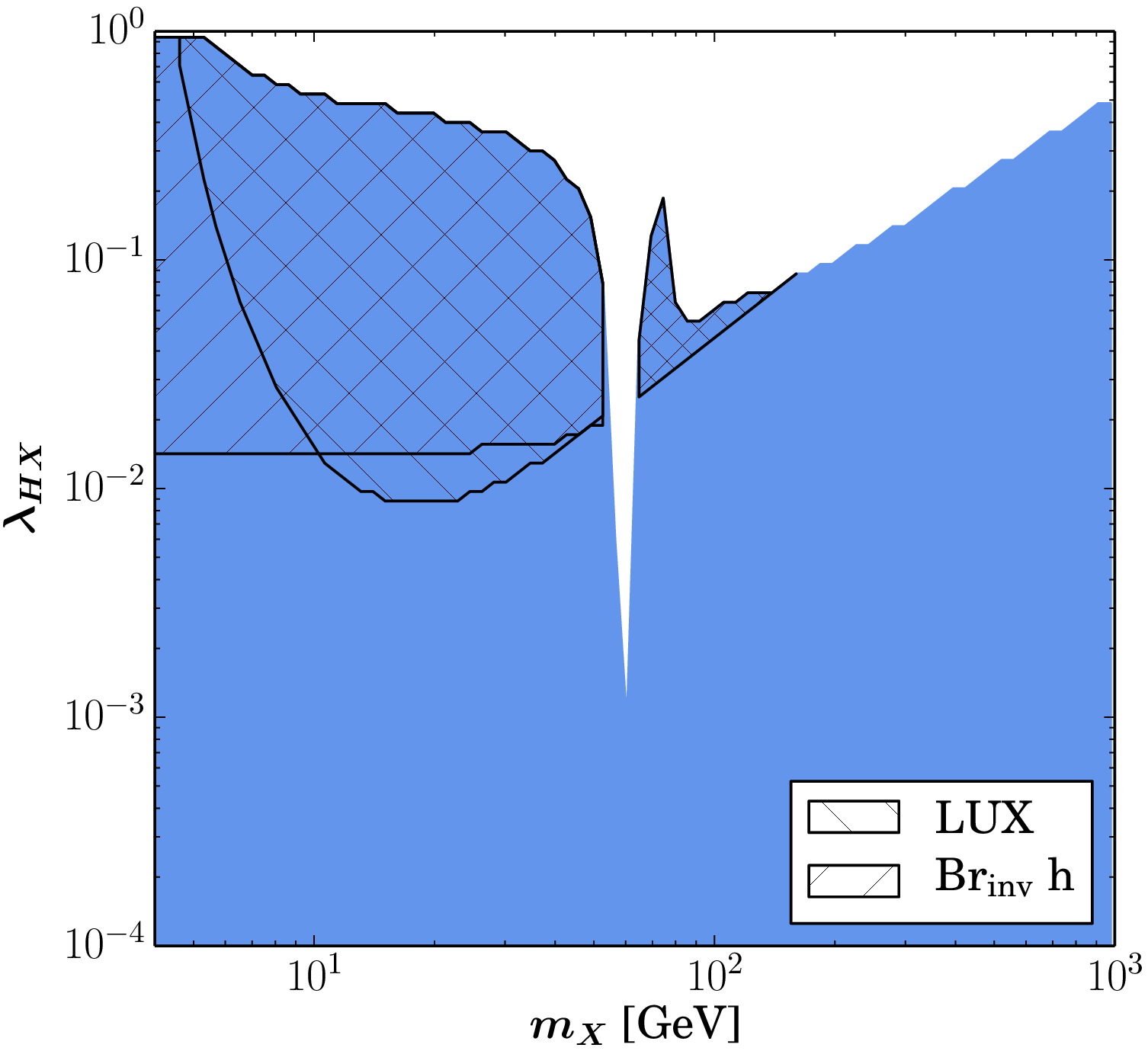}
\includegraphics[width=7.6cm]{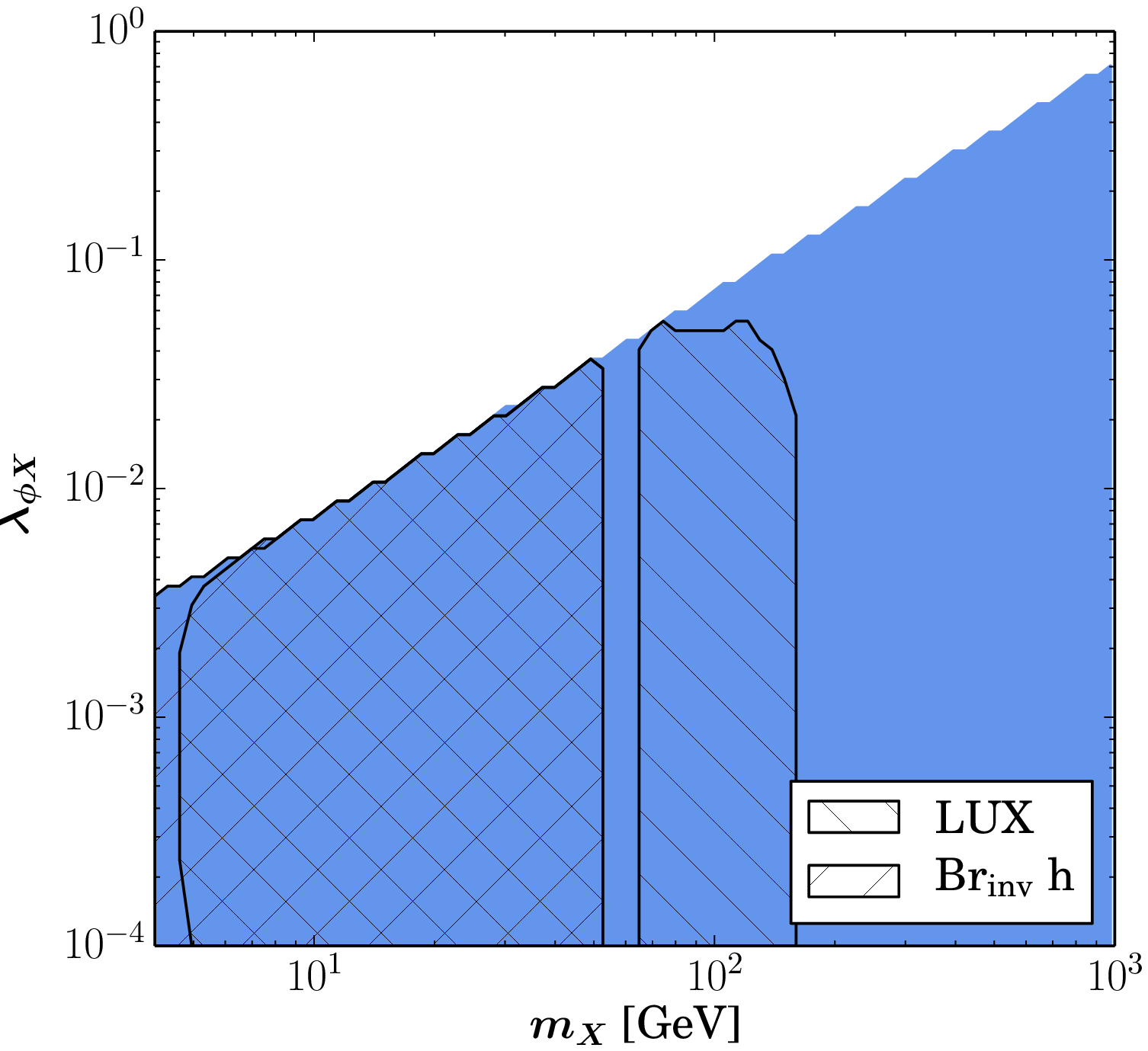}
\end{center}
\vspace{-0.8cm}
\caption{\sl \textbf{\textit{Self-annihilation scenario.}}
Same as Fig.~\ref{figdd} but marginalizing with respect to $\lfX$ ($\lHX$) in the left (right) panel.
We also include the constraint from the Higgs invisible decay width.
}
\label{figddb}
\end{figure}
Similarly, the left (right) panel of Fig.~\ref{figddb} shows in blue the regions of the parameter space that could give rise to the measures DM relic abundance in the plane $[\mx,\,\lHX]$ ($[\mx,\,\lfX]$) after marginalizing with respect to $\lfX$ ($\lHX$); that is, each plot includes all the possible values of the corresponding quartic coupling.
The empty regions, corresponding to high values for the couplings, always generate a DM underabundance.
In addition, both the LUX and the Higgs invisible decay exclusions are shown in the figure as hatched regions.
Again, those areas correspond to the points that could be excluded for some combinations of the parameters.
The unhatched regions are always allowed.

\subsection{Semi-annihilating Scenario}

\begin{figure}[t]
\begin{center}
{
\unitlength=1.0 pt
\SetScale{1.0}
\SetWidth{0.7}      
\scriptsize    
{} \qquad\allowbreak
\begin{picture}(96,38)(0,0)
\DashArrowLine(24.0,23.0)(48.0,11.0){1.0} 
\Text(24.0,23.0)[r]{$~X$}
\DashArrowLine(24.0,-1.0)(48.0,11.0){1.0} 
\Text(24.0,-1.0)[r]{$~X$}
\DashArrowLine(72.0,35.0)(48.0,11.0){1.0} 
\Text(72.0,35.0)[l]{$\bar X$}
\Text(72.0,38.0)[l]{$~$}
\DashLine(48.0,11.0)(72.0,11.0){1.0}
\Text(72.0,11.0)[l]{$h,\,\rho,\,\eta$}
\end{picture} \ 
{} \qquad\allowbreak
\begin{picture}(96,38)(0,0)
\DashArrowLine(12.0,35.0)(36.0,23.0){1.0} 
\Text(12.0,35.0)[r]{$~X$}
\DashArrowLine(12.0,11.0)(36.0,23.0){1.0} 
\Text(12.0,11.0)[r]{$~X$}
\DashArrowLine(60.0,23.0)(36.0,23.0){1.0} 
\Text(49.0,26.0)[b]{$X$}
\Text(49.0,29.0)[b]{$~$}
\DashArrowLine(84.0,35.0)(60.0,23.0){1.0} 
\Text(84.0,35.0)[l]{$\bar X$}
\Text(84.0,38.0)[l]{$~$}
\DashLine(60.0,23.0)(84.0,11.0){1.0}
\Text(84.0,11.0)[l]{$h,\,\rho$}
\end{picture} \ 
{} \qquad\allowbreak
\begin{picture}(96,38)(0,0)
\DashArrowLine(24.0,35.0)(48.0,35.0){1.0} 
\Text(24.0,35.0)[r]{$~X$}
\DashArrowLine(48.0,35.0)(48.0,11.0){1.0} 
\Text(51.0,24.0)[l]{$X$}
\Text(51.0,27.0)[l]{$~$}
\DashLine(48.0,35.0)(72.0,35.0){1.0}
\Text(72.0,35.0)[l]{$h,\,\rho$}
\DashArrowLine(24.0,11.0)(48.0,11.0){1.0} 
\Text(24.0,11.0)[r]{$~X$}
\DashArrowLine(72.0,11.0)(48.0,11.0){1.0} 
\Text(72.0,11.0)[l]{$\bar X$}
\Text(72.0,14.0)[l]{$~$}
\end{picture} \ 
}
\end{center}
\vspace{-0.8cm}
\caption{\sl \textbf{\textit{Tree level diagrams for the semi-annihilation process $\boldsymbol{X\,X\to\bar X\,S}$}}
where $S$ could be $h$, $\rho$ or $\eta$.}
\label{semiann}
\end{figure}
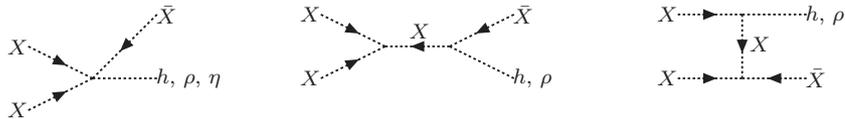

In the previous section we assumed that $\lt$ was negligible with respect to $\lHX$ and $\lfX$.
In general this may not be the case and therefore the semi-annihilation and the $3\to 2$ process have to be taken into account.
Although the latter process can still dominate, here we simply assume that it is subdominant and work under that hypothesis.
In section~\ref{3to2} we will address the general case and discuss under which circumstances the $3\to 2$ reaction can be safely ignored.
Under these conditions, the generation of relic abundance occurs not only via the self-annihilation of DM particles but also via the semi-annihilation processes DM~DM~$\to$~DM~$S$, where $S$ could be $h$, $\rho$ or $\eta$.
The corresponding Feynman diagrams at tree-level are shown in Fig.~\ref{semiann}.
Notice that the semi-annihilation into GBs is always possible and can only be suppressed by taking $\lt$ very small.

\begin{figure}[t]
\begin{center}
\includegraphics[width=8.7cm]{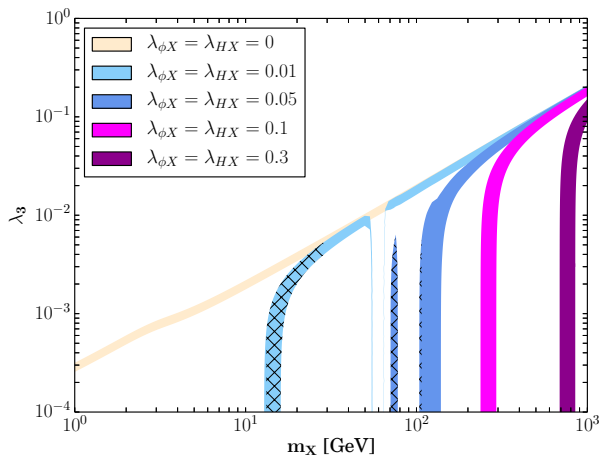}
\end{center}
\vspace{-0.8cm}
\caption{\sl \textbf{\textit{Semi-annihilation scenario.}}
Parameter space giving rise to the measured DM relic abundance in the plane $[\mx,\,\lt]$ for different combinations of the couplings $\lfX=\lHX$.
We have chosen $\mr=5$~GeV, $\sin\theta=10^{-5}$ and $\lf=0.1$.
The hatched regions are ruled out by the LUX experiment.
}
\label{figdd-l3}
\end{figure}

When $\lt$ is much greater than $\lfX$ and $\lHX$, self-annihilations can be neglected as well as the second and third diagrams of the semi-annihilation process in Fig.~\ref{semiann}; in other words, only the contact interaction diagram contributes.
Hence, the relic abundance scales like $\Omega_\text{DM}h^2\sim 1/\langle\sigma v\rangle_\text{semi}\sim\left(\mx/\lt\right)^2$.
This behavior is illustrated in Fig.~\ref{figdd-l3}, where we depict the regions of the parameter space $[\mx,\,\lt]$ giving rise to the observed relic density when $\lfX=\lHX=0$.
There we also show the equivalent regions when we relax this assumption, while still keeping $\lfX=\lHX$ for the sake of simplicity.
In this plot we choose $\mr=5$~GeV, $\sin\theta=10^{-5}$ and $\lf=0.1$.
In the case where $\lfX=\lHX$ are sizable, both the Higgs boson funnel and the kinematic openings of the annihilations into $W^+W^-$, $ZZ$ and $hh$ are visible.
Similarly, the LUX experiment rules out parts of the parameter space corresponding to DM masses ${\cal O}(10)$~GeV.
These parts are shown in the Fig.~\ref{figdd-l3} as hatched regions.

\begin{figure}[t]
\begin{center}
\includegraphics[width=8.7cm]{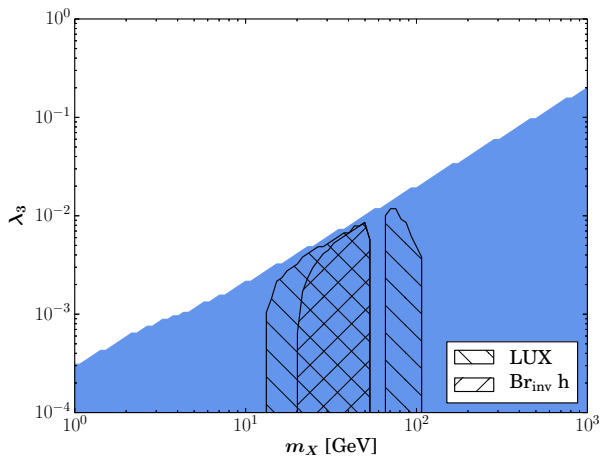}
\end{center}
\vspace{-0.8cm}
\caption{\sl \textbf{\textit{Semi-annihilation scenario.}}
Same as Fig.~\ref{figdd-l3} but marginalizing with respect to $\lfX$ and $\lHX$.
We also include the constraint from the Higgs invisible decay width.
}
\label{figddb-l3}
\end{figure}

As we did for the self-annihilation case, in Fig.~\ref{figddb-l3} we show the region of the parameter space that could give rise to the measures DM relic abundance in the plane $[\mx,\,\lt]$ after marginalizing with respect to $\lHX$ and $\lfX$.
The white region, corresponding to large values for $\lt$, always generates a DM underabundance.
Moreover, assuming that the cross-section for the semi-annihilation into a GB is never greater than about $\langle\sigma v\rangle_\text{semi}\sim 3\cdot 10^{-26}$~cm$^3$/s, this translates into the upper bound
\begin{equation}
\label{eqltmx}
\lt\lesssim 3\cdot 10^{-4}\,\left(\frac{\mx}{1\,\text{GeV}}\right)\,,
\end{equation}
which can be seen in Fig.~\ref{figddb-l3}.
In addition, we also show both the LUX and the Higgs invisible decay exclusion limits.

\section{Strongly Interacting Dark Matter}
\label{sec:SIMP}

\subsection{Dark Matter Self-interactions}
In spite of the great success of the $\Lambda$CDM scenario in explaining the formation and evolution
of cosmic structures, there are some problems at small scales (see, {\it e.g.}~\cite{Weinberg:2013aya}).
Simulations of this scenario result in cuspy 
density profile of halos towards its center and a large
number of small satellite halos. These results are challenged by 
observations. Although the inclusion of baryonic effects
can ameliorate the discrepancies, it is still unclear whether
it is necessary to change the physics of the DM sector. For instance,
self-interacting DM with a strength~\cite{Spergel:1999mh,Wandelt:2000ad,Vogelsberger:2012ku,Rocha:2012jg,Peter:2012jh,Zavala:2012us,Vogelsberger:2014pda,Elbert:2014bma}:
\begin{equation}
\label{eqsidm}
0.1\text{ cm}^2/\text{g}\lesssim\left.\frac{\sigma_\text{scatter}}{\mx}\right|_\text{obs}\lesssim10\text{ cm}^2/\text{g}
\end{equation}
can help in solving the observed discrepancies.
Yet, the Bullet Cluster~\cite{Clowe:2003tk,Markevitch:2003at,Randall:2007ph} and recent analysis of the constraints from halo shapes~\cite{Rocha:2012jg,Zavala:2012us,Peter:2012jh}, favor the following upper bound on the DM self-interacting cross section (at velocities greater than 300 km/s)
\begin{equation}
\label{eqsidm2}
\frac{\sigma_\text{scatter}}{\mx}\lesssim 1\text{ cm}^2/\text{g}\,.
\end{equation}

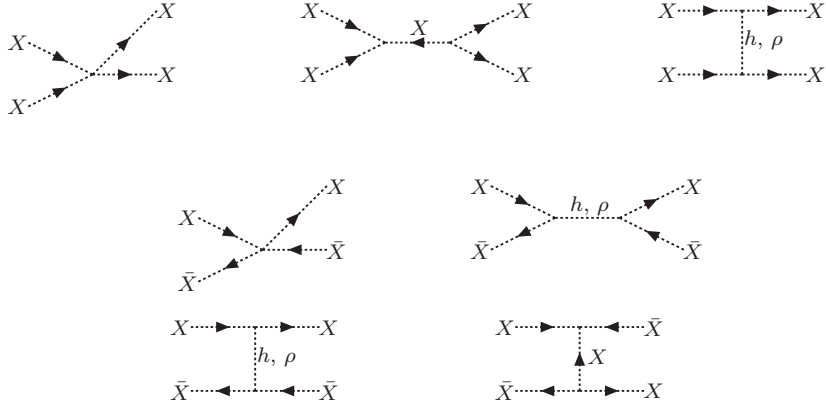
\begin{figure}[t]
\begin{center}
{
\unitlength=1.0 pt
\SetScale{1.0}
\SetWidth{0.7}      
\scriptsize    
{} \qquad\allowbreak
\begin{picture}(96,38)(0,0)
\DashArrowLine(24.0,23.0)(48.0,11.0){1.0} 
\Text(24.0,23.0)[r]{$~X$}
\DashArrowLine(24.0,-1.0)(48.0,11.0){1.0} 
\Text(24.0,-1.0)[r]{$~X$}
\DashArrowLine(48.0,11.0)(72.0,35.0){1.0} 
\Text(72.0,35.0)[l]{$X$}
\Text(72.0,38.0)[l]{$~$}
\DashArrowLine(48.0,11.0)(72.0,11.0){1.0} 
\Text(72.0,11.0)[l]{$X$}
\Text(72.0,14.0)[l]{$~$}
\end{picture} \ 
{} \qquad\allowbreak
\begin{picture}(96,38)(0,0)
\DashArrowLine(12.0,35.0)(36.0,23.0){1.0} 
\Text(12.0,35.0)[r]{$~X$}
\DashArrowLine(12.0,11.0)(36.0,23.0){1.0} 
\Text(12.0,11.0)[r]{$~X$}
\DashArrowLine(60.0,23.0)(36.0,23.0){1.0} 
\Text(49.0,26.0)[b]{$X$}
\Text(49.0,29.0)[b]{$~$}
\DashArrowLine(60.0,23.0)(84.0,35.0){1.0} 
\Text(84.0,35.0)[l]{$X$}
\Text(84.0,38.0)[l]{$~$}
\DashArrowLine(60.0,23.0)(84.0,11.0){1.0} 
\Text(84.0,11.0)[l]{$X$}
\Text(84.0,14.0)[l]{$~$}
\end{picture} \ 
{} \qquad\allowbreak
\begin{picture}(96,38)(0,0)
\DashArrowLine(24.0,35.0)(48.0,35.0){1.0} 
\Text(24.0,35.0)[r]{$~X$}
\DashLine(48.0,35.0)(48.0,11.0){1.0}
\Text(49.0,24.0)[l]{$h,\,\rho$}
\DashArrowLine(48.0,35.0)(72.0,35.0){1.0} 
\Text(72.0,35.0)[l]{$X$}
\Text(72.0,38.0)[l]{$~$}
\DashArrowLine(24.0,11.0)(48.0,11.0){1.0} 
\Text(24.0,11.0)[r]{$~X$}
\DashArrowLine(48.0,11.0)(72.0,11.0){1.0} 
\Text(72.0,11.0)[l]{$X$}
\Text(72.0,14.0)[l]{$~$}
\end{picture} \ 
}
{
\unitlength=1.0 pt
\SetScale{1.0}
\SetWidth{0.7}      
\scriptsize    
{} \qquad\allowbreak
\begin{picture}(96,38)(0,0)
\DashArrowLine(24.0,23.0)(48.0,11.0){1.0} 
\Text(24.0,23.0)[r]{$~X$}
\DashArrowLine(48.0,11.0)(24.0,-1.0){1.0} 
\Text(24.0,-1.0)[r]{$~\bar X$}
\DashArrowLine(48.0,11.0)(72.0,35.0){1.0} 
\Text(72.0,35.0)[l]{$X$}
\Text(72.0,38.0)[l]{$~$}
\DashArrowLine(72.0,11.0)(48.0,11.0){1.0} 
\Text(72.0,11.0)[l]{$\bar X$}
\Text(72.0,14.0)[l]{$~$}
\end{picture} \ 
{} \qquad\allowbreak
\begin{picture}(96,38)(0,0)
\DashArrowLine(12.0,35.0)(36.0,23.0){1.0} 
\Text(12.0,35.0)[r]{$~X$}
\DashArrowLine(36.0,23.0)(12.0,11.0){1.0} 
\Text(12.0,11.0)[r]{$~\bar X$}
\DashLine(36.0,23.0)(60.0,23.0){1.0}
\Text(49.0,24.0)[b]{$h,\,\rho$}
\DashArrowLine(60.0,23.0)(84.0,35.0){1.0} 
\Text(84.0,35.0)[l]{$X$}
\Text(84.0,38.0)[l]{$~$}
\DashArrowLine(84.0,11.0)(60.0,23.0){1.0} 
\Text(84.0,11.0)[l]{$\bar X$}
\Text(84.0,14.0)[l]{$~$}
\end{picture} \ 
\\\vspace{0.5cm}
}

{
\unitlength=1.0 pt
\SetScale{1.0}
\SetWidth{0.7}      
\scriptsize    
{} \qquad\allowbreak
\begin{picture}(96,38)(0,0)
\DashArrowLine(24.0,35.0)(48.0,35.0){1.0} 
\Text(24.0,35.0)[r]{$~X$}
\DashLine(48.0,35.0)(48.0,11.0){1.0}
\Text(49.0,24.0)[l]{$h,\,\rho$}
\DashArrowLine(48.0,35.0)(72.0,35.0){1.0} 
\Text(72.0,35.0)[l]{$X$}
\Text(72.0,38.0)[l]{$~$}
\DashArrowLine(48.0,11.0)(24.0,11.0){1.0} 
\Text(24.0,11.0)[r]{$~\bar X$}
\DashArrowLine(72.0,11.0)(48.0,11.0){1.0} 
\Text(72.0,11.0)[l]{$\bar X$}
\Text(72.0,14.0)[l]{$~$}
\end{picture} \ 
{} \qquad\allowbreak
\begin{picture}(96,38)(0,0)
\DashArrowLine(24.0,35.0)(48.0,35.0){1.0} 
\Text(24.0,35.0)[r]{$~X$}
\DashArrowLine(48.0,11.0)(48.0,35.0){1.0} 
\Text(51.0,24.0)[l]{$X$}
\Text(51.0,27.0)[l]{$~$}
\DashArrowLine(72.0,35.0)(48.0,35.0){1.0} 
\Text(72.0,35.0)[l]{$\bar X$}
\Text(72.0,38.0)[l]{$~$}
\DashArrowLine(48.0,11.0)(24.0,11.0){1.0} 
\Text(24.0,11.0)[r]{$~\bar X$}
\DashArrowLine(48.0,11.0)(72.0,11.0){1.0} 
\Text(72.0,11.0)[l]{$X$}
\Text(72.0,14.0)[l]{$~$}
\end{picture} \ 
}
\end{center}
\vspace{-0.8cm}
\caption{\sl \textbf{\textit{Diagrams for the elastic scattering process $\boldsymbol{X\,X\to X\,X}$ and $\boldsymbol{X\,\bar{X}\to X\,\bar{X}}$.}}
}
\label{figxxb-xxb}
\end{figure}

The corresponding elastic scattering processes in our model are shown in Fig.~\ref{figxxb-xxb}.
From the diagrams it is possible to see that the exchange of a $\rho$ or a $h$ boson lighter than the DM can possibly lead to non-perturbative effects such as long range interactions.
In this section we limit ourselves to the perturbative scenario and we therefore assume $\mr$, $\mh\gg\mx$.
A posteriori we will find that this is also necessary for the DM production to take place via the $3\to 2$ annihilation process (see section~\ref{3to2}).
Under this assumption
the elastic scattering cross-section in the non-relativistic limit reads
\begin{equation}
\sigma_\text{scatter}=\frac{3\,(4\,\lX^2 - 20\,\lX\,\Z^2 + 57\,\Z^4)}{64\pi\,\mx^2},
\label{som}
\end{equation}
where $\Z\equiv\lt\,\vevf/\mx$.
The $\Z$ parameter can not be arbitrarily large.
In fact, perturbative unitarity requires the $s$-wave amplitude of the process $X\bar X\to X\bar X$ to be bounded from above~\cite{Kilian:2003yw}.
Considering $\mr$ and $\mh\gg\mx$, this translate into
\begin{equation}
2\lX + 9\Z^2 + \left[\frac{\lHX\,\vevH\,\sin\theta+\lfX\,\vevf\,\cos\theta}{\mr}\right]^2 + \left[\frac{\lHX\,\vevH\,\cos\theta-\lfX\,\vevf\,\sin\theta}{\mh}\right]^2 < 4\pi\,,
\label{pertunifull}
\end{equation}
which implies the following weaker bounds
\begin{equation}
\Z < \frac23\sqrt{\pi}\sim 1.2\qquad\text{and}\qquad\lX < 2\pi\,.
\label{pertunipar}
\end{equation}
The process $X X\to X X$ also sets an upper bound on the same parameters, however it is never as stringent as the one of Eq.~\eqref{pertunifull}.

\begin{figure}[t]
\begin{center}
\includegraphics[width=8.7cm]{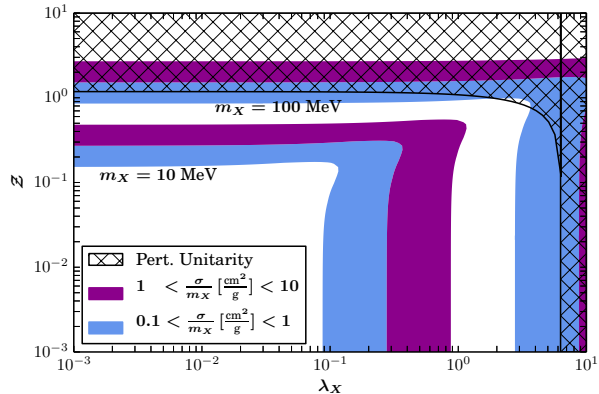}
\end{center}
\vspace{-0.8cm}
\caption{\sl \textbf{\textit{Self-interacting cross-section.}}
Regions of the parameter space preferred by the astrophysical observations of small scale structures anomalies, for different values of the DM mass (see Eq.~\eqref{eqsidm}).
The areas disfavored by the Bullet Cluster ({\it i.e.} Eq.~\eqref{eqsidm2}) are also shown.
In the hatched region perturbative unitarity is lost.
}
\label{figsidm}
\end{figure}

Fig.~\ref{figsidm} shows the regions of the plane $[\lX,\,\Z]$ fulfilling Eq.~\eqref{eqsidm} for different values of the DM mass.
In addition, we depict the areas disfavored by the Bullet Cluster  for those masses ({\it i.e.} Eq.~\eqref{eqsidm2}) and we also hatch the regions where perturbative unitarity is lost.
In this figure a limit on the DM mass is manifest, corresponding to
\begin{equation}
\mx \lesssim 173\,\text{MeV}\,,
\label{limitmx1}
\end{equation}
which is the mass that saturates the perturbative unitary bound of \eqref{pertunipar} and that is still in agreement with the astrophysical lower bound of Eq.~\eqref{eqsidm}.

\subsection[$3\to 2$ Dark Matter Scenario]{$\boldsymbol{3\to 2}$ Dark Matter Scenario}\label{3to2}

Instead of the usual $2\to 2$ annihilation scenarios described in previous sections, another mechanism for the reduction of the comoving DM number density is the annihilation via $3\to 2$ processes.
Here we discuss this case, assuming that it is the dominant process for the relic density computation, 
or equivalently that $\gamma_\text{self}$, $\gamma_\text{semi}\ll\gamma_{3\to 2}$ at the epoch of the freeze-out.

As explained in the Appendix~\ref{AppB}, it is necessary for the DM to release the kinetic energy produced in the $3\to2$ reactions. This can be fulfilled if DM is in kinetic equilibrium with either SM particles or the GBs. The first possibility is realized by means of DM scattering off electrons. However, it requires a large Higgs portal. 
This can be understood from the effective Lagrangian between electrons and DM. Assuming $m_h,\, \mr \gg \mx$ and $\theta \ll 1$, we have  
\begin{equation}
{\cal L} = \frac{\lHX\, m_e }{m_h^2}  X^* X \,\bar{e}{e}\,,
\end{equation}
and one can write
\begin{equation}
\langle \sigma v \rangle_\text{kin} =\frac{\epsilon^2}{\mx^2}\,,\hspace{30pt}\text{with}\hspace{10pt}
\epsilon=\frac{\lHX\, m_e\, \mx}{\sqrt{8\pi}\,\mh^2}\,.
\end{equation}
where $m_e$ is the electron mass.
In order to keep the kinetic equilibrium between the DM and the electrons from this interaction, it is necessary that  $ \epsilon \gtrsim 5\times10^{-9} $~\cite{Hochberg:2014dra}, or equivalently, that $ \lHX \gtrsim 1.5\times 10^3 \,(m_e/\mx)$, which is too large if Eq.~\eqref{limitmx1} holds. 

As a result, the only possibility is to have kinetic  equilibrium between the DM with the GBs. The corresponding rate depends on $\lfX$ and for $m_h,\, \mr \gg \mx$ is mostly driven by a contact interaction (see Fig~\ref{selfann}). Because of this, it is not suppressed by a heavy scale in contrast to the previous case. Moreover, although this scattering rate should be greater than the Hubble parameter at the epoch of the freeze-out, we must insure that the rate  for the self-annihilation process into GBs -which is obtained by crossing symmetry- is negligible. This is possible because $\Gamma_\text{kin}/\Gamma_\text{ann} \sim n_\eta/n_\text{DM} \gg 1$ at the freeze-out, due to the sub-dominance of $n_\text{DM}$ during the radiation-dominated epoch. We conclude that $\lfX$ must be sizable but not too large in order to avoid self-annihilations into GBs.  

\begin{figure}[t]
\begin{center}
{
\unitlength=1.0 pt
\SetScale{1.0}
\SetWidth{0.7}      
\scriptsize    
{} \qquad\allowbreak
\begin{picture}(120,62)(0,0)
\DashArrowLine(24.0,47.0)(48.0,47.0){1.0} 
\Text(24.0,47.0)[r]{$~X$}
\DashArrowLine(48.0,47.0)(48.0,11.0){1.0} 
\Text(51.0,30.0)[l]{$X$}
\Text(51.0,33.0)[l]{$~$}
\DashLine(48.0,47.0)(72.0,47.0){1.0}
\Text(61.0,48.0)[b]{$h,\,\rho$}
\DashArrowLine(24.0,11.0)(48.0,11.0){1.0} 
\Text(24.0,11.0)[r]{$~X$}
\DashArrowLine(96.0,11.0)(48.0,11.0){1.0} 
\Text(96.0,11.0)[l]{$\bar X$}
\Text(96.0,14.0)[l]{$~$}
\DashArrowLine(72.0,47.0)(96.0,59.0){1.0} 
\Text(96.0,59.0)[l]{$X$}
\Text(96.0,62.0)[l]{$~$}
\DashArrowLine(96.0,35.0)(72.0,47.0){1.0} 
\Text(96.0,35.0)[l]{$\bar X$}
\Text(96.0,38.0)[l]{$~$}
\end{picture} \ 
{} \qquad\allowbreak
\begin{picture}(120,62)(0,0)
\DashArrowLine(24.0,47.0)(48.0,47.0){1.0} 
\Text(24.0,47.0)[r]{$~X$}
\DashLine(48.0,47.0)(48.0,11.0){1.0}
\Text(49.0,30.0)[l]{$h,\,\rho$}
\DashArrowLine(48.0,47.0)(72.0,47.0){1.0} 
\Text(61.0,50.0)[b]{$X$}
\Text(61.0,53.0)[b]{$~$}
\DashArrowLine(24.0,11.0)(48.0,11.0){1.0} 
\Text(24.0,11.0)[r]{$~X$}
\DashArrowLine(48.0,11.0)(96.0,11.0){1.0} 
\Text(96.0,11.0)[l]{$X$}
\Text(96.0,14.0)[l]{$~$}
\DashArrowLine(96.0,59.0)(72.0,47.0){1.0} 
\Text(96.0,59.0)[l]{$\bar X$}
\Text(96.0,62.0)[l]{$~$}
\DashArrowLine(96.0,35.0)(72.0,47.0){1.0} 
\Text(96.0,35.0)[l]{$\bar X$}
\Text(96.0,38.0)[l]{$~$}
\end{picture} \ 
{} \qquad\allowbreak
\begin{picture}(120,62)(0,0)
\DashArrowLine(36.0,59.0)(60.0,59.0){1.0} 
\Text(36.0,59.0)[r]{$~X$}
\DashLine(60.0,59.0)(60.0,35.0){1.0}
\Text(61.0,48.0)[l]{$h,\,\rho$}
\DashArrowLine(60.0,59.0)(84.0,59.0){1.0} 
\Text(84.0,59.0)[l]{$X$}
\Text(84.0,62.0)[l]{$~$}
\DashArrowLine(60.0,35.0)(60.0,11.0){1.0} 
\Text(63.0,24.0)[l]{$X$}
\Text(63.0,27.0)[l]{$~$}
\DashArrowLine(84.0,35.0)(60.0,35.0){1.0} 
\Text(84.0,35.0)[l]{$\bar X$}
\Text(84.0,38.0)[l]{$~$}
\DashArrowLine(36.0,11.0)(60.0,11.0){1.0} 
\Text(36.0,11.0)[r]{$~X$}
\DashArrowLine(84.0,11.0)(60.0,11.0){1.0} 
\Text(84.0,11.0)[l]{$\bar X$}
\Text(84.0,14.0)[l]{$~$}
\end{picture} \ 
{} \qquad\allowbreak
\begin{picture}(120,62)(0,0)
\DashArrowLine(36.0,59.0)(60.0,59.0){1.0} 
\Text(36.0,59.0)[r]{$~X$}
\DashLine(60.0,59.0)(60.0,11.0){1.0}
\Text(61.0,36.0)[l]{$h,\,\rho$}
\DashArrowLine(60.0,59.0)(84.0,59.0){1.0} 
\Text(84.0,59.0)[l]{$X$}
\Text(84.0,62.0)[l]{$~$}
\DashArrowLine(36.0,11.0)(60.0,11.0){1.0} 
\Text(36.0,11.0)[r]{$~X$}
\DashArrowLine(84.0,35.0)(60.0,11.0){1.0} 
\Text(84.0,35.0)[l]{$\bar X$}
\Text(84.0,38.0)[l]{$~$}
\DashArrowLine(84.0,11.0)(60.0,11.0){1.0} 
\Text(84.0,11.0)[l]{$\bar X$}
\Text(84.0,14.0)[l]{$~$}
\end{picture} \ 
{} \qquad\allowbreak
\begin{picture}(120,62)(0,0)
\DashArrowLine(24.0,59.0)(48.0,47.0){1.0} 
\Text(24.0,59.0)[r]{$~X$}
\DashArrowLine(24.0,35.0)(48.0,47.0){1.0} 
\Text(24.0,35.0)[r]{$~X$}
\DashArrowLine(72.0,47.0)(48.0,47.0){1.0} 
\Text(61.0,50.0)[b]{$X$}
\Text(61.0,53.0)[b]{$~$}
\DashArrowLine(96.0,59.0)(72.0,47.0){1.0} 
\Text(96.0,59.0)[l]{$\bar X$}
\Text(96.0,62.0)[l]{$~$}
\DashLine(72.0,47.0)(72.0,23.0){1.0}
\Text(73.0,36.0)[l]{$h,\,\rho$}
\DashArrowLine(72.0,23.0)(96.0,35.0){1.0} 
\Text(96.0,35.0)[l]{$X$}
\Text(96.0,38.0)[l]{$~$}
\DashArrowLine(96.0,11.0)(72.0,23.0){1.0} 
\Text(96.0,11.0)[l]{$\bar X$}
\Text(96.0,14.0)[l]{$~$}
\end{picture} \ 
{} \qquad\allowbreak
\begin{picture}(120,62)(0,0)
\DashArrowLine(24.0,35.0)(48.0,23.0){1.0} 
\Text(24.0,35.0)[r]{$~X$}
\DashArrowLine(24.0,11.0)(48.0,23.0){1.0} 
\Text(24.0,11.0)[r]{$~X$}
\DashArrowLine(96.0,59.0)(48.0,23.0){1.0} 
\Text(96.0,59.0)[l]{$\bar X$}
\Text(96.0,62.0)[l]{$~$}
\DashLine(48.0,23.0)(72.0,23.0){1.0}
\Text(61.0,24.0)[b]{$h,\,\rho$}
\DashArrowLine(72.0,23.0)(96.0,35.0){1.0} 
\Text(96.0,35.0)[l]{$X$}
\Text(96.0,38.0)[l]{$~$}
\DashArrowLine(96.0,11.0)(72.0,23.0){1.0} 
\Text(96.0,11.0)[l]{$\bar X$}
\Text(96.0,14.0)[l]{$~$}
\end{picture} \ 
{} \qquad\allowbreak
\begin{picture}(120,62)(0,0)
\DashArrowLine(24.0,35.0)(48.0,23.0){1.0} 
\Text(24.0,35.0)[r]{$~X$}
\DashArrowLine(24.0,11.0)(48.0,23.0){1.0} 
\Text(24.0,11.0)[r]{$~X$}
\DashArrowLine(48.0,23.0)(96.0,59.0){1.0} 
\Text(96.0,59.0)[l]{$X$}
\Text(96.0,62.0)[l]{$~$}
\DashArrowLine(48.0,23.0)(72.0,23.0){1.0} 
\Text(61.0,26.0)[b]{$X$}
\Text(61.0,29.0)[b]{$~$}
\DashArrowLine(96.0,35.0)(72.0,23.0){1.0} 
\Text(96.0,35.0)[l]{$\bar X$}
\Text(96.0,38.0)[l]{$~$}
\DashArrowLine(96.0,11.0)(72.0,23.0){1.0} 
\Text(96.0,11.0)[l]{$\bar X$}
\Text(96.0,14.0)[l]{$~$}
\end{picture} \ 
{} \qquad\allowbreak
\begin{picture}(120,62)(0,0)
\DashArrowLine(36.0,35.0)(60.0,35.0){1.0} 
\Text(36.0,35.0)[r]{$~X$}
\DashArrowLine(60.0,35.0)(60.0,11.0){1.0} 
\Text(63.0,24.0)[l]{$X$}
\Text(63.0,27.0)[l]{$~$}
\DashArrowLine(60.0,35.0)(84.0,59.0){1.0} 
\Text(84.0,59.0)[l]{$X$}
\Text(84.0,62.0)[l]{$~$}
\DashArrowLine(84.0,35.0)(60.0,35.0){1.0} 
\Text(84.0,35.0)[l]{$\bar X$}
\Text(84.0,38.0)[l]{$~$}
\DashArrowLine(36.0,11.0)(60.0,11.0){1.0} 
\Text(36.0,11.0)[r]{$~X$}
\DashArrowLine(84.0,11.0)(60.0,11.0){1.0} 
\Text(84.0,11.0)[l]{$\bar X$}
\Text(84.0,14.0)[l]{$~$}
\end{picture} \ 
{} \qquad\allowbreak
\begin{picture}(120,62)(0,0)
\DashArrowLine(24.0,47.0)(48.0,35.0){1.0} 
\Text(24.0,47.0)[r]{$~X$}
\DashArrowLine(24.0,23.0)(48.0,35.0){1.0} 
\Text(24.0,23.0)[r]{$~X$}
\DashArrowLine(72.0,35.0)(48.0,35.0){1.0} 
\Text(61.0,38.0)[b]{$X$}
\Text(61.0,41.0)[b]{$~$}
\DashArrowLine(72.0,35.0)(96.0,59.0){1.0} 
\Text(96.0,59.0)[l]{$X$}
\Text(96.0,62.0)[l]{$~$}
\DashArrowLine(96.0,35.0)(72.0,35.0){1.0} 
\Text(96.0,35.0)[l]{$\bar X$}
\Text(96.0,38.0)[l]{$~$}
\DashArrowLine(96.0,11.0)(72.0,35.0){1.0} 
\Text(96.0,11.0)[l]{$\bar X$}
\Text(96.0,14.0)[l]{$~$}
\end{picture} \ 
{} \qquad\allowbreak
\begin{picture}(120,62)(0,0)
\DashArrowLine(24.0,59.0)(48.0,47.0){1.0} 
\Text(24.0,59.0)[r]{$~X$}
\DashArrowLine(24.0,35.0)(48.0,47.0){1.0} 
\Text(24.0,35.0)[r]{$~X$}
\DashArrowLine(72.0,47.0)(48.0,47.0){1.0} 
\Text(61.0,50.0)[b]{$X$}
\Text(61.0,53.0)[b]{$~$}
\DashArrowLine(72.0,47.0)(96.0,59.0){1.0} 
\Text(96.0,59.0)[l]{$X$}
\Text(96.0,62.0)[l]{$~$}
\DashArrowLine(72.0,47.0)(72.0,23.0){1.0} 
\Text(75.0,36.0)[l]{$X$}
\Text(75.0,39.0)[l]{$~$}
\DashArrowLine(96.0,35.0)(72.0,23.0){1.0} 
\Text(96.0,35.0)[l]{$\bar X$}
\Text(96.0,38.0)[l]{$~$}
\DashArrowLine(96.0,11.0)(72.0,23.0){1.0} 
\Text(96.0,11.0)[l]{$\bar X$}
\Text(96.0,14.0)[l]{$~$}
\end{picture} \ 
{} \qquad\allowbreak
\begin{picture}(120,62)(0,0)
\DashArrowLine(36.0,59.0)(60.0,59.0){1.0} 
\Text(36.0,59.0)[r]{$~X$}
\DashArrowLine(60.0,35.0)(60.0,59.0){1.0} 
\Text(63.0,48.0)[l]{$X$}
\Text(63.0,51.0)[l]{$~$}
\DashArrowLine(84.0,59.0)(60.0,59.0){1.0} 
\Text(84.0,59.0)[l]{$\bar X$}
\Text(84.0,62.0)[l]{$~$}
\DashArrowLine(60.0,35.0)(60.0,11.0){1.0} 
\Text(63.0,24.0)[l]{$X$}
\Text(63.0,27.0)[l]{$~$}
\DashArrowLine(60.0,35.0)(84.0,35.0){1.0} 
\Text(84.0,35.0)[l]{$X$}
\Text(84.0,38.0)[l]{$~$}
\DashArrowLine(36.0,11.0)(60.0,11.0){1.0} 
\Text(36.0,11.0)[r]{$~X$}
\DashArrowLine(84.0,11.0)(60.0,11.0){1.0} 
\Text(84.0,11.0)[l]{$\bar X$}
\Text(84.0,14.0)[l]{$~$}
\end{picture} \ 
}
\end{center}
\vspace{-0.8cm}
\caption{\sl \textbf{\textit{Tree level diagrams for the annihilation $\boldsymbol{X\,X\leftrightarrow X\,\bar X\,\bar X}$.}}
The first two rows contain processes that are mediated via the exchange of $h$, $\rho$ and $X$ whereas the last two rows involve solely DM particles.}
\label{threetwoa}
\end{figure}
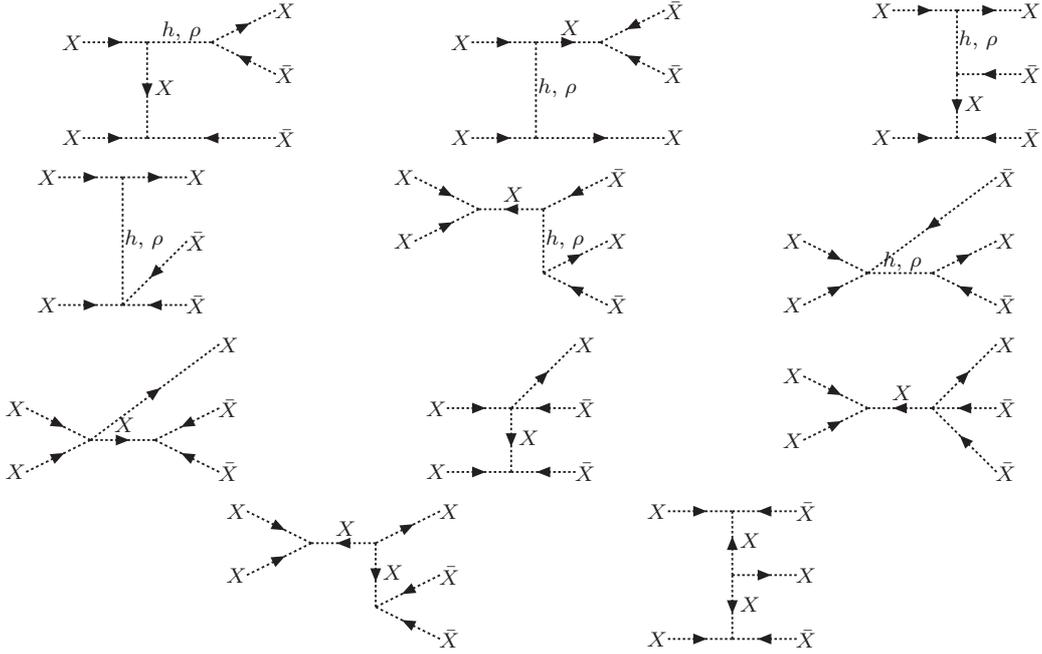

\begin{figure}[t]
\begin{center}
{
\unitlength=1.0 pt
\SetScale{1.0}
\SetWidth{0.7}      
\scriptsize    
{} \qquad\allowbreak
\begin{picture}(120,62)(0,0)
\DashArrowLine(36.0,59.0)(60.0,59.0){1.0} 
\Text(36.0,59.0)[r]{$~X$}
\DashLine(60.0,59.0)(60.0,11.0){1.0}
\Text(61.0,36.0)[l]{$h,\,\rho$}
\DashArrowLine(60.0,59.0)(84.0,59.0){1.0} 
\Text(84.0,59.0)[l]{$X$}
\Text(84.0,62.0)[l]{$~$}
\DashArrowLine(60.0,11.0)(36.0,11.0){1.0} 
\Text(36.0,11.0)[r]{$~\bar X$}
\DashArrowLine(60.0,11.0)(84.0,35.0){1.0} 
\Text(84.0,35.0)[l]{$X$}
\Text(84.0,38.0)[l]{$~$}
\DashArrowLine(60.0,11.0)(84.0,11.0){1.0} 
\Text(84.0,11.0)[l]{$X$}
\Text(84.0,14.0)[l]{$~$}
\end{picture} \ 
{} \qquad\allowbreak
\begin{picture}(120,62)(0,0)
\DashArrowLine(24.0,59.0)(48.0,59.0){1.0} 
\Text(24.0,59.0)[r]{$~X$}
\DashLine(48.0,59.0)(48.0,23.0){1.0}
\Text(49.0,42.0)[l]{$h,\,\rho$}
\DashArrowLine(48.0,59.0)(96.0,59.0){1.0} 
\Text(96.0,59.0)[l]{$X$}
\Text(96.0,62.0)[l]{$~$}
\DashArrowLine(48.0,23.0)(24.0,23.0){1.0} 
\Text(24.0,23.0)[r]{$~\bar X$}
\DashArrowLine(72.0,23.0)(48.0,23.0){1.0} 
\Text(61.0,26.0)[b]{$X$}
\Text(61.0,29.0)[b]{$~$}
\DashArrowLine(72.0,23.0)(96.0,35.0){1.0} 
\Text(96.0,35.0)[l]{$X$}
\Text(96.0,38.0)[l]{$~$}
\DashArrowLine(72.0,23.0)(96.0,11.0){1.0} 
\Text(96.0,11.0)[l]{$X$}
\Text(96.0,14.0)[l]{$~$}
\end{picture} \ 
{} \qquad\allowbreak
\begin{picture}(120,62)(0,0)
\DashArrowLine(36.0,59.0)(60.0,59.0){1.0} 
\Text(36.0,59.0)[r]{$~X$}
\DashLine(60.0,59.0)(60.0,35.0){1.0}
\Text(61.0,48.0)[l]{$h,\,\rho$}
\DashArrowLine(60.0,59.0)(84.0,59.0){1.0} 
\Text(84.0,59.0)[l]{$X$}
\Text(84.0,62.0)[l]{$~$}
\DashArrowLine(60.0,11.0)(60.0,35.0){1.0} 
\Text(63.0,24.0)[l]{$X$}
\Text(63.0,27.0)[l]{$~$}
\DashArrowLine(60.0,35.0)(84.0,35.0){1.0} 
\Text(84.0,35.0)[l]{$X$}
\Text(84.0,38.0)[l]{$~$}
\DashArrowLine(60.0,11.0)(36.0,11.0){1.0} 
\Text(36.0,11.0)[r]{$~\bar X$}
\DashArrowLine(60.0,11.0)(84.0,11.0){1.0} 
\Text(84.0,11.0)[l]{$X$}
\Text(84.0,14.0)[l]{$~$}
\end{picture} \ 
{} \qquad\allowbreak
\begin{picture}(120,62)(0,0)
\DashArrowLine(24.0,47.0)(48.0,35.0){1.0} 
\Text(24.0,47.0)[r]{$~X$}
\DashArrowLine(48.0,35.0)(24.0,23.0){1.0} 
\Text(24.0,23.0)[r]{$~\bar X$}
\DashLine(48.0,35.0)(72.0,35.0){1.0}
\Text(61.0,36.0)[b]{$h,\,\rho$}
\DashArrowLine(72.0,35.0)(96.0,59.0){1.0} 
\Text(96.0,59.0)[l]{$X$}
\Text(96.0,62.0)[l]{$~$}
\DashArrowLine(72.0,35.0)(96.0,35.0){1.0} 
\Text(96.0,35.0)[l]{$X$}
\Text(96.0,38.0)[l]{$~$}
\DashArrowLine(72.0,35.0)(96.0,11.0){1.0} 
\Text(96.0,11.0)[l]{$X$}
\Text(96.0,14.0)[l]{$~$}
\end{picture} \ 
{} \qquad\allowbreak
\begin{picture}(120,62)(0,0)
\DashArrowLine(24.0,59.0)(48.0,47.0){1.0} 
\Text(24.0,59.0)[r]{$~X$}
\DashArrowLine(48.0,47.0)(24.0,35.0){1.0} 
\Text(24.0,35.0)[r]{$~\bar X$}
\DashLine(48.0,47.0)(72.0,47.0){1.0}
\Text(61.0,48.0)[b]{$h,\,\rho$}
\DashArrowLine(72.0,47.0)(96.0,59.0){1.0} 
\Text(96.0,59.0)[l]{$X$}
\Text(96.0,62.0)[l]{$~$}
\DashArrowLine(72.0,23.0)(72.0,47.0){1.0} 
\Text(75.0,36.0)[l]{$X$}
\Text(75.0,39.0)[l]{$~$}
\DashArrowLine(72.0,23.0)(96.0,35.0){1.0} 
\Text(96.0,35.0)[l]{$X$}
\Text(96.0,38.0)[l]{$~$}
\DashArrowLine(72.0,23.0)(96.0,11.0){1.0} 
\Text(96.0,11.0)[l]{$X$}
\Text(96.0,14.0)[l]{$~$}
\end{picture} \ 
{} \qquad\allowbreak\\
\begin{picture}(120,62)(0,0)
\DashArrowLine(24.0,35.0)(48.0,23.0){1.0} 
\Text(24.0,35.0)[r]{$~X$}
\DashArrowLine(48.0,23.0)(24.0,11.0){1.0} 
\Text(24.0,11.0)[r]{$~\bar X$}
\DashArrowLine(48.0,23.0)(96.0,59.0){1.0} 
\Text(96.0,59.0)[l]{$X$}
\Text(96.0,62.0)[l]{$~$}
\DashArrowLine(72.0,23.0)(48.0,23.0){1.0} 
\Text(61.0,26.0)[b]{$X$}
\Text(61.0,29.0)[b]{$~$}
\DashArrowLine(72.0,23.0)(96.0,35.0){1.0} 
\Text(96.0,35.0)[l]{$X$}
\Text(96.0,38.0)[l]{$~$}
\DashArrowLine(72.0,23.0)(96.0,11.0){1.0} 
\Text(96.0,11.0)[l]{$X$}
\Text(96.0,14.0)[l]{$~$}
\end{picture} \ 
{} \qquad\allowbreak
\begin{picture}(120,62)(0,0)
\DashArrowLine(36.0,35.0)(60.0,35.0){1.0} 
\Text(36.0,35.0)[r]{$~X$}
\DashArrowLine(60.0,11.0)(60.0,35.0){1.0} 
\Text(63.0,24.0)[l]{$X$}
\Text(63.0,27.0)[l]{$~$}
\DashArrowLine(60.0,35.0)(84.0,59.0){1.0} 
\Text(84.0,59.0)[l]{$X$}
\Text(84.0,62.0)[l]{$~$}
\DashArrowLine(60.0,35.0)(84.0,35.0){1.0} 
\Text(84.0,35.0)[l]{$X$}
\Text(84.0,38.0)[l]{$~$}
\DashArrowLine(60.0,11.0)(36.0,11.0){1.0} 
\Text(36.0,11.0)[r]{$~\bar X$}
\DashArrowLine(60.0,11.0)(84.0,11.0){1.0} 
\Text(84.0,11.0)[l]{$X$}
\Text(84.0,14.0)[l]{$~$}
\end{picture} \ 
{} \qquad\allowbreak
\begin{picture}(120,62)(0,0)
\DashArrowLine(24.0,47.0)(48.0,47.0){1.0} 
\Text(24.0,47.0)[r]{$~X$}
\DashArrowLine(48.0,11.0)(48.0,47.0){1.0} 
\Text(51.0,30.0)[l]{$X$}
\Text(51.0,33.0)[l]{$~$}
\DashArrowLine(72.0,47.0)(48.0,47.0){1.0} 
\Text(61.0,50.0)[b]{$X$}
\Text(61.0,53.0)[b]{$~$}
\DashArrowLine(48.0,11.0)(24.0,11.0){1.0} 
\Text(24.0,11.0)[r]{$~\bar X$}
\DashArrowLine(48.0,11.0)(96.0,11.0){1.0} 
\Text(96.0,11.0)[l]{$X$}
\Text(96.0,14.0)[l]{$~$}
\DashArrowLine(72.0,47.0)(96.0,59.0){1.0} 
\Text(96.0,59.0)[l]{$X$}
\Text(96.0,62.0)[l]{$~$}
\DashArrowLine(72.0,47.0)(96.0,35.0){1.0} 
\Text(96.0,35.0)[l]{$X$}
\Text(96.0,38.0)[l]{$~$}
\end{picture} \ 
}
\end{center}
\vspace{-0.8cm}
\caption{\sl \textbf{\textit{Tree level diagrams for the annihilation $\boldsymbol{X\,\bar X\leftrightarrow X\,X\,X}$.}}
}
\label{threetwob}
\end{figure}
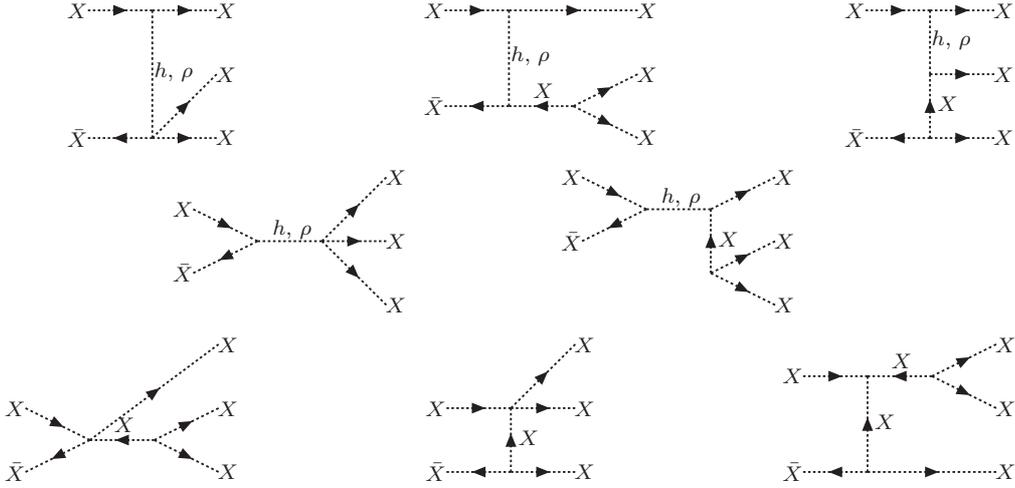

In section~\ref{secdr}, it was implicitly assumed that the GB did not heat up after its decoupling from the SM plasma.
In fact, this is not the case here because DM is not in kinetic equilibrium with the SM and transfers all its entropy exclusively to the GBs when it annihilates via the $3\to 2$ mechanism.
As a result, Eq.~\eqref{NeffdefG} must be modified accordingly in order to account for this effect and thus $T^d_\eta$ must be larger than the muon mass so that the contribution to $N_\text{eff}$ is negligeable.
This always takes place here because we consider $\mh,\,\mr\gg\mx\sim 100$~MeV, which implies that the processes establishing the equilibrium are very suppressed for temperatures around the muon mass.

With this in mind, we can consider the $3\to2$ reactions in this model. All of them are equivalent, up to $CP$ conjugation, to either $\bar X\,\bar X\,X\to X\,X$ or $X\,X\,X\to X\,\bar X$.
In Figs.~\ref{threetwoa} and~\ref{threetwob} we show the corresponding diagrams at tree-level.
The corresponding  cross-sections can be calculated analytically if we work in the non-relativistic approximation (see Appendix~\ref{AppC}).
In fact, the resulting expressions can be further simplified in the case $\gamma_\text{self}\ll\gamma_{3\to 2}$ because of the following reason.
As for self-annihilation, some of the $3\to 2$ diagrams are proportional to $\lHX$ and $\lfX$.
In fact, this happens for those in the first and second rows of Figs.~\ref{threetwoa} and~\ref{threetwob} simply because they involve exchange of $\rho$ or $h$ particles.
In contrast, the diagrams of the third and fourth lines depend uniquely on different couplings, namely $\lt$ and $\lX$.
Consequently, the contribution of the first set of diagrams is presumably negligible, if self-annihilations are subdominant.
The fact that $\mr$ and $\mh\gg\mx$ reinforces this argument.
However, notice that for this to be true it is not necessary that $\lfX$ vanishes. 
Under these assumptions we find
\begin{equation}\label{sv32}
\langle\sigma v^2\rangle_{3\to 2}=\frac{\sqrt{5}\Z^2}{768\pi\,\mx^5}\left[4\left(2\lambda_X+ 9\Z^2\right)^2 + \frac{1}{32}\left(74\lambda_X-117\Z^2\right)^2\right]+\mathcal{O}\left(\lHX\,,\lfX\right)^2\,,
\end{equation}
where $\Z = \lt\,\vevf/\mx$ was already introduced.
Because self- and semi-annihilations into GBs are always kinematically allowed, Eq.~\eqref{sv32} must be compared with the cross-sections associated to these processes
\begin{equation}
\langle\sigma v\rangle_\text{self}=\mathcal{O}\left(\lHX\,,\lfX\right)^2\,,\qquad\langle\sigma v\rangle_\text{semi}=\frac{27\,\lt^2}{256\pi\,\mx^2}\,.
\end{equation}
We already discussed the first process. Regarding the second one, 
although $\langle\sigma v\rangle_\text{semi}$ and $\langle\sigma v^2\rangle_{3\to 2}$ have different mass dimensions and can not be compared directly, it is clear that the semi-annihilations are suppressed for small $\lt$, and that the $3\to 2$ is important when $\Z$ is large.
Moreover, a suppression in the $3\to 2$ process due to $\lt$ can be compensated with a large $\vevf$ with respect to the DM mass. Under these circumstances we can make sure that $\gamma_\text{self}$, $\gamma_\text{semi}\ll\gamma_{3\to 2}$, while still having kinetic equilibrium between the GBs and the DM. We remark that the DM mass is not required to be in the GeV range because we assume no equilibrium between the GBs and SM plasma and consequently there is no significant contribution to $N_\text{eff}$. 

In order to study the regions of the parameter space where the $3\to 2$ reproduces the observed relic abundance, instead of solving numerically the Boltzmann equations directly we work in the freeze-out approximation.

\subsection[The $3\to 2$ Freeze-out Approximation]{The $\boldsymbol{3\to2}$ Freeze-out Approximation}

When the temperature of the plasma is much greater than the DM mass, the Boltzmann Eq.~\eqref{be} has  the solution $n= n_\text{eq}$, which corresponds to the chemical equilibrium. However, once the temperature drops below the DM mass, the rate for the process $2 \to 3$ is kinematically suppressed compared to the $3\to 2$ reaction. As a result the DM number density deviates from its equilibrium value. In fact, it eventually becomes much greater than its equilibrium value $  n \gg n_\text{eq}$. Hence, for these temperatures and when there are no self- and semi-annihilations, Eq.~\eqref{be} can be approximated by
\begin{eqnarray}
\frac{dn}{dt} + 3\,H\,n \approx - n^3 \langle \sigma v^2\rangle_{3 \to 2}.
\label{beFO}
\end{eqnarray}

When the DM particles are non-relativistic, the cross-section $\langle \sigma v^2 \rangle$ is independent of the temperature. 
In that case, using standard methods, Eq.~\eqref{beFO} admits the following solution 
\begin{equation}
\langle \sigma v^2 \rangle_{3\to2} = \left( 8.65\, \text{GeV}^{-5}\right) \,x_\text{FO}^4 \, g_\text{FO}^{-1.5} \left(\frac{\mx}{1\, \text{GeV}} \right)^{-2}\,,
\label{FOcondition}
\end{equation}
where FO stands for the freeze-out, $g$ are the number of relativistic degrees of freedom and $x=\mx/T$. Here $T$ is the DM temperature which equals that of the GB plasma.  
It is remarkable that, in contrast to the standard self-annihilation scenario, here the cross-section that matches the observed relic density depends on the DM mass.  In order to estimate $x_\text{FO}$, it is necessary to establish when  the annihilation rate per particle $n_\text{eq}^2 \langle \sigma v \rangle_{3 \to 2}$ drops below the expansion rate of the Universe. Using Eq.~\eqref{FOcondition} it is found that this happens when freeze-out temperature satisfies
\begin{equation}
x_\text{FO} = 20.6 + \log \left[\left(\frac{\mx}{100 \,\text{MeV}}\right)\left(\frac{g_\text{FO}}{10.75}\right)^{-1}\left(\frac{x_\text{FO}}{20.6}\right)^{1.5}\right]\,.
\label{x0}
\end{equation}
Because the second term of this equation depends logarithmically  on the DM mass, it is a very good approximation to take $x_\text{FO} \approx 20$.

It is somewhat interesting to calculate the value of $\langle \sigma v \rangle_{2\to3}$ associated to Eq.~\eqref{FOcondition}. To that end, notice that at the freeze-out the expansion rate is  $H = n_\text{eq}\, \langle \sigma v \rangle_{2\to 3 } =  n_\text{eq}^2 \langle \sigma v^2 \rangle_{3\to 2}$ and as a result $n_\text{eq} = \sqrt{H/ \langle \sigma v^2 \rangle_{3\to2}}$. According to Eq.~\eqref{32to23} we obtain
\begin{eqnarray}
\langle \sigma v \rangle_{2\to3} &=& \sqrt{H\,\langle \sigma v^2 \rangle_{3\to2}} = \left( 1.08\,\times 10^{-9}\,\text{GeV}^{-2} \right) \,x_\text{FO} \, g_\text{FO}^{-0.5}\nonumber\\
                                 &\approx& 5.9 \times 10^{-26}\text{ cm}^3/\text{s}\,.
\end{eqnarray}
In contrast to Eq.~\eqref{FOcondition}, notice that this equation is independent of the DM mass and is similar to the usual condition for the WIMP scenario.\\

\begin{figure}[t]
\begin{center}
\includegraphics[width=8.7cm]{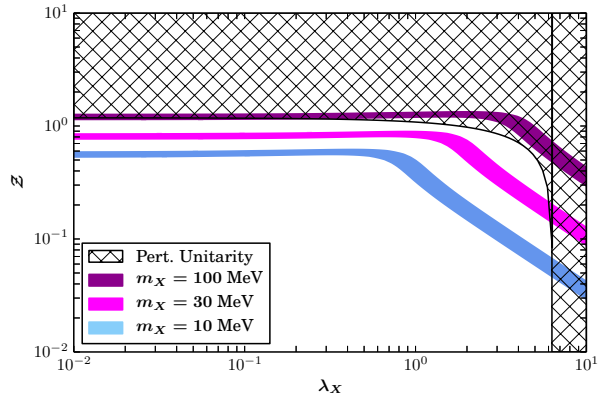}
\end{center}
\vspace{-0.8cm}
\caption{\sl \textbf{\textit{Freeze-out approximation for the $\boldsymbol{3\to 2}$ scenario.}}
Parameter space giving rise to the measured DM relic abundance for different DM masses.
In the hatched region perturbative unitarity is lost.
}
\label{figdm32}
\end{figure}

We use this formalism in order to calculate the relic density. In Fig.~\ref{figdm32} we show the regions of the plane $[\lX,\,\Z]$ in agreement with the observed relic density
for different values of the DM mass when the $3\to2$ process dominates the DM production. 
They are obtained using Eqs.~\eqref{sv32} and~\eqref{FOcondition}, taking into account an uncertainty of $25\%$ in order to account for the error 
in the freeze-out and the non-relativistic approximations.
As apparent from the figure, there are two regimes.
On the one hand, for small values of $\lX$ the DM abundance is set by $\Z$.
On the other hand, when $\lX$ is of order one both variables are relevant for the DM dynamics.
In particular notice that $\Z$ can not vanish.
Fig.~\ref{figdm32} also includes the region where perturbative unitarity is lost, which implies an upper limit on the DM mass as shown in the figure:
\begin{equation}
\mx\lesssim115~\text{MeV}\,.
\end{equation}
Let us note that this limit, obtained from the relic density, is comparable to the one in Eq.~\eqref{limitmx1} coming from demanding significant DM self-interactions.

\subsection[Self-interactions and the $3\to 2$ Mechanism]{Self-interactions and the $\boldsymbol{3\to 2}$ Mechanism}

\begin{figure}[t]
\begin{center}
\includegraphics[width=7.6cm]{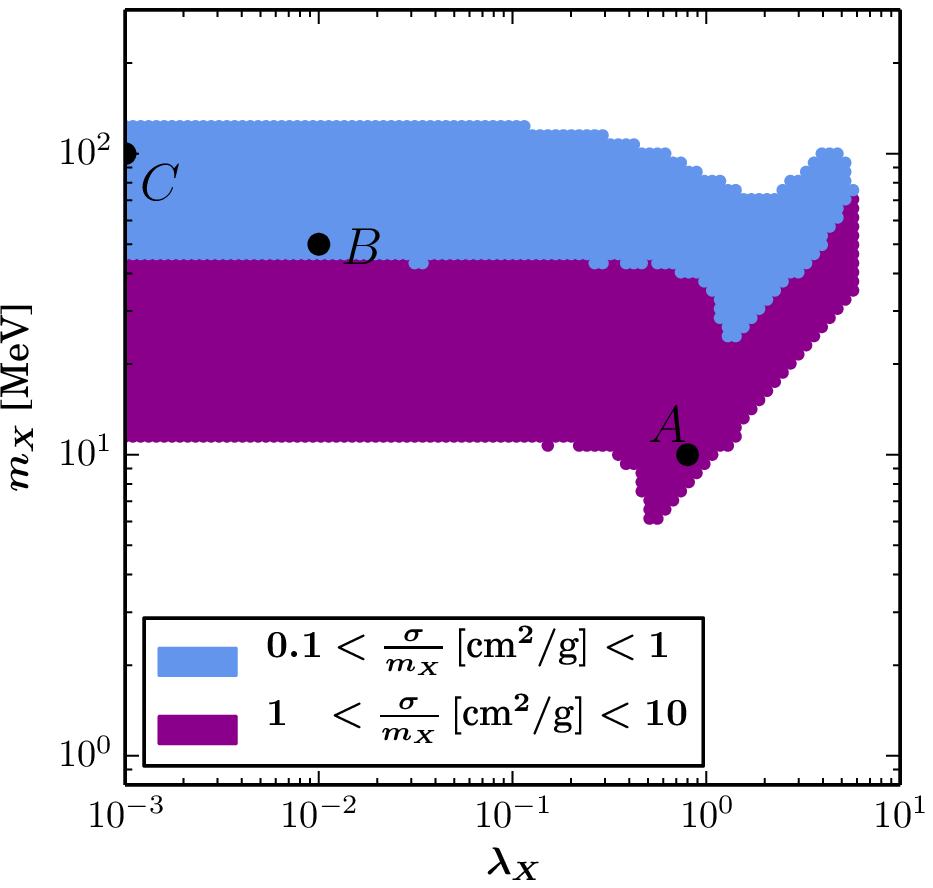}
\includegraphics[width=7.6cm]{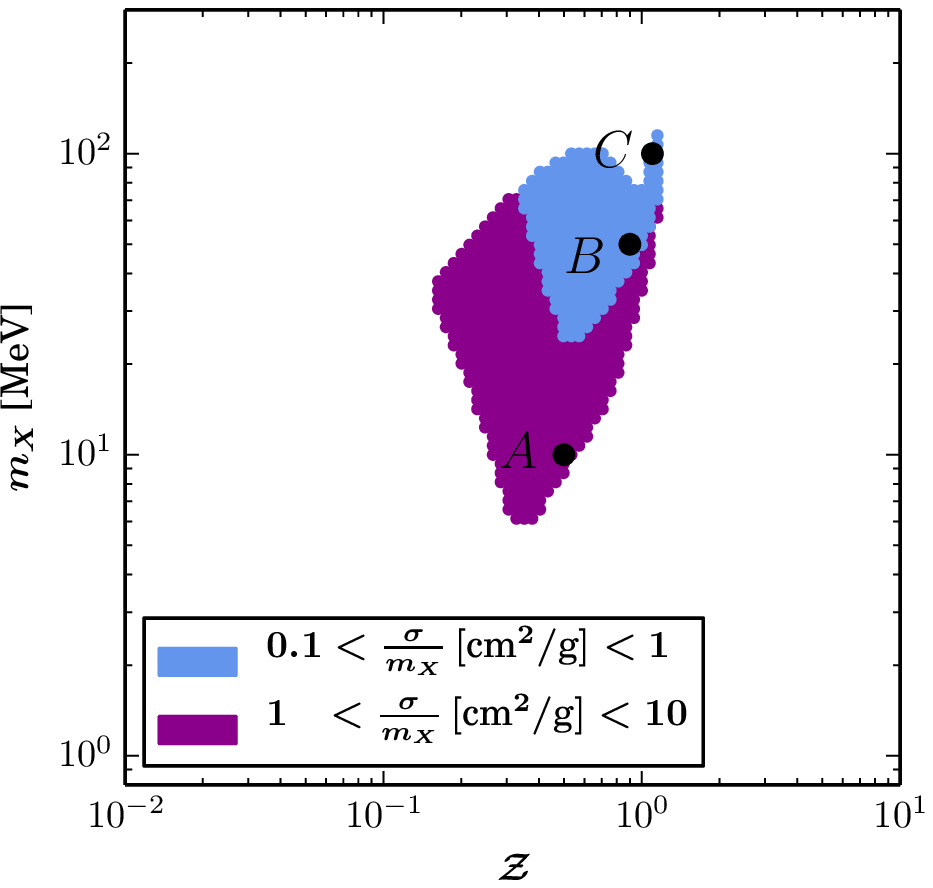}
\end{center}
\vspace{-0.8cm}
\caption{\sl \textbf{\textit{Self-interacting dark matter and the $\boldsymbol{3\to 2}$ mechanism.}}
Parameter space giving rise to the measured DM relic abundance, preferred by the astrophysical observations of small scale structures anomalies and satisfying perturbative unitarity.
The three benchmark points defined in Table~\ref{tabbenchmarks} are also shown.
}
\label{som23}
\end{figure}

It is interesting that the variables that control DM self-interactions also determine the $3\to 2$ process, as shown in Eqs.~\eqref{som} and~\eqref{sv32}.
This suggests that both phenomena have a common origin, which we attribute to the fact that DM is strongly interacting.
Fig.~\ref{som23} illustrates this by showing the regions of the parameter space where the relic abundance is generated via the $3\to 2$ mechanism and where the self-interaction cross-section simultaneously fulfills the astrophysical hints from small scale structures, as in Eq.~\eqref{eqsidm}.
In addition, we show the areas disfavored by the Bullet Cluster ({\it i.e.} Eq.~\eqref{eqsidm2}).
All this has been done under the assumption that perturbative unitarity is valid.
The figure conclusively shows that the DM mass is bounded both from above and below
\begin{equation}\label{desmas}
7~\text{MeV}\lesssim\mx\lesssim 115~\text{MeV}\,,
\end{equation}
which is in agreement with the findings of~\cite{Hochberg:2014dra}.
Similarly, for the $\Z$ parameter we find
\begin{equation}
0.35\lesssim\Z<\frac23\sqrt{\pi}\,.
\end{equation}

We illustrate this scenario with three benchmark points which are defined in Table~\ref{tabbenchmarks} and shown in Fig.~\ref{som23}.
They correspond to different masses in the MeV range, which reproduce simultaneously the observed relic abundance via the $3\to 2$ mechanism and the hints on the self-interaction cross-section.
Whereas points {\bf B} and {\bf C} give rise to a $\sigma_\text{scatter}/\mx$ in agreement with the Bullet Cluster observations, point {\bf A} is disfavored due to its small mass.
In every case $\lt$, $\lHX$ and $\lfX$ are chosen small enough in order to ensure that semi-annihilations and self-annihilations are subdominant (see Table~\ref{tabbenchmarks}).
Nevertheless, we require that DM is in kinetic equilibrium with the GBs via the $\lfX$ coupling, as argued before.
We again find that the scale at which the $U(1)_\text{DM}$ symmetry breaks down has to be at least in the TeV range so that $\Z$ is of order one.
We would like to remark that all these points are in agreement with collider constraints because the $\rho$ boson is very heavy and the invisible decay width of the Higgs is very small.
In fact, the most important contribution to the latter comes from the process $h\to X\bar X$ with a branching ratio of approximately $1.5\cdot 10^{-5}$.
The other channels ($\rho\rho$, $\eta\eta$, $XXX$ and $\bar X\bar X\bar X$) contribute less than $\sim 10^{-12}$.
{\renewcommand{\arraystretch}{1.5}
\begin{table}[ht]
\centering
\begin{tabular}{|c||c|c|c|c|} \hline
                                & \bf{A}           & \bf{B}           & \bf{C}           \\\hline\hline
$\boldsymbol{\mx}$ \bf{[MeV]}   & $10$             & $50$             & $100$            \\
$\boldsymbol{\lt}$              & $5\cdot 10^{-7}$ & $2\cdot 10^{-6}$ & $5\cdot 10^{-6}$ \\
$\boldsymbol{\lX}$              & $0.8$            & $10^{-2}$        & $10^{-3}$        \\
$\boldsymbol{\lf}$              & $10^{-4}$        & $10^{-4}$        & $10^{-4}$        \\
$\boldsymbol{\Z}$               & $0.5$            & $0.9$            & $1.1$            \\\hline
$\boldsymbol{\mr}$ \bf{[GeV]}   & $141$            & $318$            & $311$            \\
$\boldsymbol{\vevf}$ \bf{[TeV]} & $10$             & $22.5$           & $22$             \\\hline\hline
$\boldsymbol{\langle\sigma v\rangle_{2\to 3}}$ \bf{[cm$\boldsymbol{^{3}}$/s]} & $6.6\cdot 10^{-26}$ & $4.8\cdot 10^{-26}$ & $4.5\cdot 10^{-26}$ \\
$\boldsymbol{\gamma_\text{\bf{semi}}/\gamma_{3\to 2}}$ & $0.03$  & $0.03$ & $0.05$  \\
$\boldsymbol{\gamma_\text{\bf{self}}/\gamma_{3\to 2}}$ & $4\cdot 10^{-3}$  & $2\cdot 10^{-4}$ & $4\cdot 10^{-5}$  \\\hline
$\boldsymbol{\sigma_\text{\bf{scatter}}/\mx}$ \bf{[cm$\boldsymbol{^2}$/g]} & $6.9$ & $0.91$ & $0.27$ \\\hline
\end{tabular}
\caption{\sl\textbf{\textit{Benchmark points}} that generate the measured DM relic abundance mainly via the $3\to 2$ processes.
Here $\sin\theta=10^{-5}$ and $\lHX=10^{-4}$.
Similarly, in order to keep the DM in kinetic equilibrium until its freeze-out we set $\lfX=10^{-6}$.
$\mr$ and $\vevf$ are derived quantities.}
\label{tabbenchmarks}
\end{table}}


\section{Conclusions}\label{sec:conclusions}
We have presented a scalar extension of the Standard Model (SM) consisting of two additional complex fields with charges one and three under a global $U(1)_\text{DM}$ symmetry.
The nonzero vacuum expectation value of the latter field spontaneously breaks the global symmetry down to a remnant $\Zt$.
This leads to the appearance of a Goldstone boson (GB), two Higgs-like particles, one of which is identified with the SM scalar, and a complex field that, due to the discrete symmetry, can not decay and is therefore a candidate for dark matter (DM).

This model is constrained by different observations.
In first place, the enlargement of the scalar sector could potentially modify the dynamics of the recently found SM scalar.
In particular, its decay into invisible particles due to opening of new channels involving the GB and the DM.
Secondly, the exchange of the Higgs-like particles can mediate elastic scattering between the DM and SM particles leading to detectable signals in direct detection experiments for DM.
Lastly, GB can mimic neutrinos in the Cosmic Microwave Background, significantly altering the so-called effective number of neutrinos $N_\text{eff}$.
We have found that these constraints are always satisfied if the Higgs-like particles are heavier than $\sim 5$~GeV, and if their mixing angle is sufficiently small, namely smaller than about $10^{-5}$.

In these regions of the parameter space we have investigated the thermal production of DM.
On the one hand, DM can self-annihilate either into SM particles by means of the Higgs portal, or into the non-DM additional scalars via the GB portal.
Similarly, DM can semi-annihilate, that is two DM particles can be converted into another DM particle and a non-DM scalar.
Because the GB is massless, both annihilation processes into GBs are always kinematically open and they therefore play an important role in the DM production.
For DM in the GeV range, the typical values for the Higgs and GB portals as well as the coupling responsible for semi-annihilations are similar to the weak couplings of the SM.
Because of this reason, when one of these two processes dominates, DM behaves as a weakly interacting massive particle (WIMP).

On the other hand, our model provides for a novel DM production mechanism, the so-called $3 \to 2$ mechanism consisting of the annihilation of three DM particles into two of them,
which is possible due to the $\Zt$ symmetry. We find regions in the parameter where this mechanism is the dominant one.
By studying the freeze-out approximation of the corresponding Boltzmann equation, and assuming the absence of non-perturbative effects, 
we have found that only DM masses in the MeV range can reproduce the observed relic abundance.
In addition, for this mechanism to dominate over the other production modes, the Higgs and GB portals are required to be suppressed.
Also, it is found that the $U(1)_\text{DM}$ symmetry must break down at least at the TeV scale.
This naturally leads to relatively strong cubic DM self-interactions.
Because of this, this scenario is a realization of the strongly interacting massive particle (SIMP) paradigm.
Remarkably, the same mass range and similar DM cubic couplings naturally give rise to self-interacting cross-sections which could solve 
the discrepancies between observations and simulations of small-scale structures in the Universe, as shown in Eq.~\eqref{eqsidm}.
In fact, by performing a dedicated analysis of the parameter space we have found that in this scenario the DM mass must be in between 7 and 115~MeV.
In order to illustrate our findings, we have studied three benchmark points (see Table~\ref{tabbenchmarks}), 
compatible with all experimental constraints, in which the relic abundance is produced via the $3\to2$ mechanism and 
where the DM self-interactions are in the range suggested by astrophysics.

Direct and indirect detection experiments as well as collider searches will continue closing in on the parameter space of the WIMP scenario, 
due to the comparatively large couplings with the SM particles.
Although this is more challenging for the SIMP scenario, astrophysics opens up a new window on the dynamics of DM self-interactions.
This is the case, for instance, of the benchmark point {\bf A} which is in tension with observations of the Bullet Cluster and recent analysis of the constraints of halo shapes.
Consequently, new astrophysical observations can shed light on the SIMP scenario, potentially constraining further its parameter space.

\acknowledgments
NB is supported by the São Paulo Research Foundation (FAPESP) under grants 2011/11973-4 and 2013/01792-8.
CGC is supported by the IISN and the Belgian Federal Science Policy through the Interuniversity Attraction Pole P7/37 “Fundamental Interactions”. RR is partially supported by a FAPESP grant 2011/11973-4 and by a CNPq research grant. The authors would like to thank Eduardo Pontón and  Gary Steigman for valuable discussions.
Also to Alexander Pukhov for providing a MicrOMEGAs routine that calculates the $2\to 3$ thermal averaged cross sections and Tim Stefaniak for proving the limits from OPAL. 


\appendix
\newpage
\section{The Boltzmann Equation}
\label{AppB}

In this Appendix we consider the Boltzmann equation for the DM particle and how to calculate the corresponding annihilation rates.  Since we consider no asymmetry between $X$ and $\bar{X}$,  we can think of the $\Zt$ charge as an internal degree of freedom for the DM. As a result the DM density is given by 

\begin{equation}
n(t) = 2 \int \frac{d^3p}{(2\pi)^3}\,f(E,t)\,
\end{equation}
where $f(E,t)$ is the DM phase-space distribution.  
The evolution of the DM density is given by the Boltzmann equation
\begin{equation}
\frac{dn}{dt}+3\,H n = - \sum_{\alpha,\beta} (N_{\alpha}- N_{\beta})(\hat\gamma_{\alpha \to \beta}-\hat\gamma_{\beta \to \alpha}) 
\end{equation}
where $H$ is the Hubble parameter, while $N_\alpha$ and $N_\beta$ are the the number of DM particles in the initial and final states.  In addition, the interaction rates are given by   
\begin{eqnarray}
\hat\gamma_{\alpha \to \beta}= 
\int \left(\prod_{i \in \alpha} \frac{d^3 p_i}{2E_i(2\pi)^3} f_i \right) \left(\prod_{j \in \beta} \frac{d^3 p_j}{2E_j(2\pi)^3}\right)
(2\pi)^4\delta^4(p_\alpha -  p_\beta) \times
 |\mathcal{M}_{\alpha\to\beta}|^2 \,,
\label{Brates}
\end{eqnarray}
where $|\mathcal{M}_{\alpha\to\beta}|^2$ is the square invariant amplitude including possible symmetry factors in both the initial and final state, respectively. Without loss of generality we assume $N_\alpha>1$. For instance, in self-annihilation processes $N_\alpha=2$ and $N_\beta=0$, in semi-annihilations $N_\alpha=2$ and $N_\beta=1$,  and for the $3\to 2$ reactions $N_\alpha=3$ and $N_\beta=2$.

We assume now that
the non-DM particles are in chemical equilibrium at a common temperature T during the period of interest. Their phase-space distribution is thus given by $f^{\text{eq}} (E,\,T) \equiv \exp({-E/T})$. Although this is not true for the DM particles,  they are assumed to be in kinetic equilibrium, as discussed below. If this is the case,  because of symmetry considerations in a Friedmann–Lemaître–Robertson–Walker Universe, their phase-space density $f(E,\,t)$ is proportional to the chemical equilibrium distribution $f^{\text{eq}}(E,\,T)$, with a proportionality factor independent of the three-momentum~\cite{Bernstein:1985th}. This assumption  implies
\begin{eqnarray}
\hat\gamma_{\alpha \to \beta}=  \left(\frac{n}{n_\text{eq}} \right)^{N_\alpha} \gamma_{\alpha \to \beta}\,,
\end{eqnarray}
where $\gamma_{\alpha \to \beta}$ corresponds to the rate of Eq.~\eqref{Brates} when $f_i=f_i^\text{eq}$.
Furthermore, due to the principle of detailed balance and $CP$ conservation $\gamma_{\alpha \to \beta}= \gamma_{\beta \to \alpha}$. Then, after adding over all the possible self-, semi- and $3\to2$ annihilation processes, the Boltzmann equation can be cast as 

\begin{equation}
\frac{dn}{dt}+3 H n=
- 2\left[ \left( \frac{n}{n_{\text{eq}}}\right)^2 -1 \right] \gamma_\text{self} -  
\frac{n}{n_{\text{eq}}}  \left[ \frac{n}{n_{\text{eq}}} -1 \right] \gamma_\text{semi} -
\left(\frac{n}{n_{\text{eq}}} \right)^2 \left[ \frac{n}{n_{\text{eq}}} -1 \right] \gamma_{3  \to 2}
\end{equation}
or as
\begin{equation}
\frac{dn}{dt}+3\,H\,n=
- \langle\sigma v  \rangle_\text{self}\left(n^2 -n_\text{eq}^2   \right) 
- \langle\sigma v  \rangle_\text{semi}\left(n^2 -n\,n_\text{eq}  \right) 
- \langle\sigma v^2\rangle_{3 \to 2}  \left(n^3 -n^2\,n_\text{eq}\right)\,, 
\end{equation}
with 
\begin{eqnarray}
\langle \sigma v \rangle_\text{self} \equiv 2\frac{\gamma_\text{self}}{n_\text{eq}^2}
\,,
\hspace{30pt}
\langle \sigma v \rangle_\text{semi} \equiv \frac{\gamma_\text{semi}}{n_\text{eq}^2} 
\hspace{15pt}
\text{and}
\hspace{15pt}
\langle \sigma v^2 \rangle_{3 \to 2} \equiv \frac{\gamma_{3 \to 2}}{n_\text{eq}^3} \,.
\end{eqnarray}

Notice that one can also define
\begin{equation}
\langle \sigma v \rangle_{2 \to 3} \equiv \frac{\gamma_{2 \to 3}}{n_\text{eq}^2} \,,
\end{equation}
which satisfies
\begin{equation}
\langle \sigma v^2 \rangle_{3 \to 2} = \frac{\langle \sigma v \rangle_{2 \to 3}}{n_\text{eq}}\,.
\label{32to23}
\end{equation}

\subsection*{Kinetic Equilibrium} 

When self or semi-annihilations are non-negligible, because of crossing symmetry, the transition amplitudes leading to them also induce scattering processes and these keep DM in kinetic equilibrium with its decaying products. In fact, when the DM particles become non-relativistic, the scattering processes occur at a faster rate than the annihilations because 
\begin{equation}
\frac{\Gamma_\text{kin}}{\Gamma_\text{ann}} = \frac{n_\text{non-DM} \langle \sigma v \rangle_\text{kin}}{n_\text{DM} \langle \sigma v \rangle_\text{ann}} \sim \frac{n_\text{non-DM}}{n_\text{DM} } \gg 1\,.
\label{KinAnnComp}
\end{equation}   
where in the last step we take relativistic non-DM particles, which are more abundant during the radiation-dominated era.  Because of this, the kinetic equilibrium is still maintained much after the DM freeze-out. 

In this work, when self and semi-annihilations can not be neglected, we further assume that all the non-DM particles are in equilibrium with each other. In particular, this means that the GB and the SM particles are in equilibrium. In order to avoid non-negligible contributions to $N_\text{eff}$, as explained in section \ref{secdr},  such equilibrium can only exist at temperatures greater than the muon mass. Because of this and because the freeze-out temperature is roughly $m_X/25 $, in this work we only consider self- and semi-annihilations in the GeV range.    

The situation is different in the opposite case, that is,  when the $3\to 2$ mechanism dominates. In this case, non-DM particles do not participate in the DM production, and therefore scattering processes between DM and GBs or SM particles do not necessarily take place. Nevertheless, the $3\to2$ processes convert a significant fraction of the mass into DM kinetic energy. If the latter is not radiated away, DM heats up altering significantly structure formation~\cite{Carlson:1992fn,Machacek:1994vg,deLaix:1995vi}~\footnote{$3\to2$ (and $4\to2$) processes are also constrained by the lifetime of the DM halo~\cite{Yamanaka:2014pva}, however this constrain is typically subdominant.}.
We can circumvent this by allowing sufficiently large DM couplings to the SM plasma or the GBs, so that the former remains in equilibrium with one of the latter~\cite{Hochberg:2014dra}. As shown in section \ref{sec:SIMP}, the first possibility requires huge couplings because the scattering rate is suppressed by the electron Yukawa coupling. Therefore, we assume the only possibility left, that is, that GBs and DM are in equilibrium. Because of Eq.~\eqref{KinAnnComp}, this is not in contradiction with the fact that self- or semi-annihilations are negligible. Notice also that in this case the DM mass is not required to be in the GeV range because we assume no equilibrium between the GBs and SM plasma.

\section{The Non-relativistic Approximation}
\label{AppC}

In order to calculate $\gamma$ for the different processes with only DM particles in the initial state, we take $f^{\text{eq}}=\exp({-E/T})$ in Eq.~\eqref{Brates}. For temperatures much smaller than the DM mass, because of the Boltzmann suppression, the integration  over the phase-space of the initial state is dominated by small three-momenta. In fact, by taking each of them equal to zero one finds
\begin{eqnarray}
\gamma_{\alpha \to \beta} \approx \left(\frac{n_\text{eq}}{4 \mx}\right)^{N_\alpha}  
\int \left(\prod_{j \in \beta} \frac{d^3 p_j}{2E_j(2\pi)^3}\right)
\Big[(2\pi)^4\delta^4(p_\alpha -  p_\beta) \times
 |\mathcal{M}_{\alpha\to\beta}|^2 \Big]_\text{NR} \,,
\label{Brateseq}
\end{eqnarray}
where the subscript NR means that all the initial state particles are taken at rest.
After performing the two-body final state phase-space integration, this implies that the cross-sections are
\begin{eqnarray}
\langle \sigma v \rangle_\text{self} &\approx& 
 \frac{1}{64 \, \pi \, \mx^2} \sum_{\alpha\,\beta}
 |\mathcal{M}_{\alpha\to\beta}|^2 \Big|_\text{NR}\,,
\\
\langle \sigma v \rangle_\text{semi} &\approx& 
\frac{3}{32\, \pi \,\mx^2} \sum_{\alpha\,\beta}
 |\mathcal{M}_{\alpha\to\beta}|^2 \Big|_\text{NR}\,,
\\
\langle \sigma v^2 \rangle_{3 \to 2} &\approx& 
\frac{\sqrt{5}}{1536\, \pi \,\mx^3}  \sum_{\alpha\,\beta}
 |\mathcal{M}_{\alpha\to\beta}|^2 \Big|_\text{NR}\,.
\end{eqnarray}
Notice that for the self- and semi-annihilation cases, these expressions are just the s-save piece of the partial wave expansion of the thermal averaged cross-sections. 

{
\renewcommand{\arraystretch}{2.0}
\begin{table}[t]
\centering
\begin{tabular}{|c|c||c|c|} \hline
$\boldsymbol{N_\alpha}$       & $\boldsymbol{\mathbb{Z}_3}$ & {\bf Process}                    & $\boldsymbol{\mx^{2(N_\alpha-2)}\, |\mathcal{M}|^2\Big|_\text{\bf NR}}$ \\\hline\hline
\multirow{2}{*}{\bf 2} &    {\bf 0}                  & $X\bar{X}\to\eta\eta$                     & 0                                     \\\cline{2-4}
                   &        {\bf 1}                  & $\bar{X}\bar{X}\to X\eta$                 & $9 \lt^2 $ \\\hline
\multirow{4}{*}{\bf 3} & \multirow{2}{*}{\bf 0}      & $XXX\to\eta\eta$                          & $\frac{8 \lf^2 \Z^2}{ (9 - 2 \lf \V^2)^2}$ \\\cline{3-4}
                       &                             & $XXX\to X\bar{X}$                         & $ 4\Z^2(2\lambda_X+ 9\Z^2)^2$ \\\cline{2-4}
                       & \multirow{2}{*}{\bf 1}      & $XX\bar{X}\to X\eta$                      & $\frac{4608\, \lt^2 \Z^2}{49  }$ \\\cline{3-4}
                       &                             & $XX\bar{X}\to\bar{X}\bar{X}$              & $\frac{\Z^2}{32}[74\lambda_X-117\Z^2]^2$  \\\hline
\multirow{6}{*}{\bf 4} & \multirow{2}{*}{\bf 0}      & $XX\bar{X}\bar{X}\to X\bar{X}$            & $\frac{1}{4 }\left[12\lX^2+63\Z^4+4\frac{9\lt^2+\Z^2(\lf(3-22\lX\V^2)-33\lX)}{3+2\lf\V^2}\right]^2$ \\\cline{3-4}
                       &                             & $XX\bar{X}\bar{X}\to\eta\eta$             & $\frac{1152\lt^4}{25}\left[\frac{\lf\V^2-3}{\lf\V^2-8}\right]^2$ \\\cline{2-4}
                       & \multirow{4}{*}{\bf 1}      & $X\bar{X}\bar{X}\bar{X}\to X\eta$         & $\frac{\lt^2}{6084}\left[182\lX-1791\Z^2\right]^2$ \\\cline{3-4}
                       &                             & $XXXX\to X\eta$                           & $\frac{2\lt^2}{3}\left[8\lX+21\Z^2\right]^2$ \\\cline{3-4}
                       &                             & $X\bar{X}\bar{X}\bar{X}\to\bar{X}\bar{X}$ & $\frac{1}{900}\left[180\lX^2-567\Z^4-4\frac{90\lt^2-729\lX\Z^2+2\lf\Z^2(81\lX\V^2-5)}{9-2\lf\V^2}\right]^2$\\\cline{3-4}
                       &                             & $XXXX\to\bar{X}\bar{X}$                   & $\frac{2 \Z^4}{75}\left[78\lX-189\Z^2+\frac{20\lf}{2\lf\V^2-9}-\frac{90\lf}{2\lf\V^2+3}\right]^2$ \\\hline
\end{tabular}
\caption{\sl \textbf{\textit{Square amplitudes in the non-relativistic limit.}}
Here $\theta,\lHX$ and $\lfX$ are neglected. We use the notation $\V = \frac{\vevf}{\mx}$ and $\Z = \lt \V$.
}
\label{table:xsections}
\end{table}
}

In Table \ref{table:xsections}, we show all the square amplitudes that contribute to these cross-sections when $\lHX$, $\lfX$ and $\theta$ are neglected.  For the sake of completeness, we also include the square amplitude for other processes with three DM particles in the initial state:  $XX\bar{X} \to X \eta$ and $XXX \to \eta \eta$. Notice that when the $3\to2$ mechanism is relevant - that is, when either $\Z$ or $\lX$ is order one and $\lt$ is small - those processes have a suppressed amplitude.  We also show some processes with four DM particles in the initial state.  Because $\lX$ and $\Z$ are bounded from perturbative unitarity considerations as implied by Eq.~\eqref{pertunifull}, those processes can not dominate over the $3\to 2$ mechanism.


\bibliographystyle{JHEP}
\bibliography{biblio}

\end{document}